\documentclass[fleqn,usenatbib]{mnras}

\usepackage{newtxtext,newtxmath}

\usepackage[T1]{fontenc}
\usepackage{ae,aecompl}

\usepackage{graphicx}	
\usepackage{amsmath}	
\usepackage{hyperref}

\usepackage{physics}
\usepackage{cleveref}
\usepackage{soul}
\usepackage[dvipsnames]{xcolor}

\newcommand{\jr}{\color{black}}

\title[Photoevaporation vs. core-powered mass-loss]{Photoevaporation vs. core-powered mass-loss: model comparison with the 3D radius gap}

\author[Rogers, J. G. et al.]{
James G. Rogers$^{1}$\thanks{E-mail: james.rogers14@imperial.ac.uk}, Akash Gupta$^{2}$, James E. Owen$^{1}$ and Hilke E. Schlichting$^{2,\,3,\,4}$
\\ 
$^{1}$Astrophysics Group, Department of Physics, Imperial College London, Prince Consort Rd, London, SW7 2AZ, UK\\
$^{2}$Department of Earth, Planetary, and Space Sciences, University of California, Los Angeles, CA 90095, USA\\
$^{3}$Department of Physics and Astronomy, University of California, Los Angeles, CA 90095, USA\\
$^{4}$Department of Earth, Atmospheric and Planetary Sciences, Massachusetts Institute of Technology, MA 02139, USA
}

\date{Accepted XXX. Received YYY; in original form ZZZ}

\pubyear{2021}

\begin{document}
\label{firstpage}
\pagerange{\pageref{firstpage}--\pageref{lastpage}}
\maketitle

\begin{abstract}
The EUV/X-ray photoevaporation and core-powered mass-loss models are both capable of reproducing the bimodality in the sizes of small, close-in exoplanets observed by the \textit{Kepler} space mission, often referred to as the ``radius gap''. However, it is unclear which of these two mechanisms dominates the atmospheric mass-loss which is likely sculpting the radius gap. In this work, we propose a new method of differentiating between the two models, which relies on analysing the radius gap in 3D parameter space. Using models for both mechanisms, and by performing synthetic transit surveys we predict the size and characteristics of a survey capable of discriminating between the two models. We find that a survey of $\gtrsim 5000$ planets, with a wide range in stellar mass and measurement uncertainties at a $\lesssim 5\%$ level is sufficient. Our methodology is robust against moderate false positive contamination of $\lesssim 10\%$. We perform our analysis on two surveys (which do not satisfy our requirements): the \textit{California Kepler Survey} and the \textit{Gaia-Kepler Survey} and find, unsurprisingly, that both data-sets are consistent with either model. We propose a hypothesis test to be performed on future surveys which can robustly ascertain which of the two mechanisms formed the radius gap, provided one dominates over the other.
\end{abstract}



\section{Introduction} \label{sec:Intro}

Of the many discoveries that the \textit{Kepler} survey revealed, perhaps the most intriguing was the abundance of small ($\lesssim$ $4R_\oplus$), low mass ($\lesssim$ $50M_\oplus$) exoplanets located close to their host star ($\lesssim$ 100 days) \citep[e.g.][]{BoruckiKeplerII,Batalha2013,Fressin2013,Petigura2013,Mullally2015,Silburt2015,Mulders2018,Zhu2018,Zink2019}. This revelation was made all the more intriguing with the detection of a bimodality in the radius distribution of these planets \citep{Fulton2017, VanEylen2018, Berger2020}. {\jr{Separated by a sparsity at $\sim 1.8R_\oplus$, ``super-Earths'' are typically $1.0$ - $1.5R_\oplus$ and have a bulk density consistent with Earth \citep[e.g.][]{Weiss2014,HaddenLithwick2014,Dressing2015,Dorn2019}. Alternatively, ``sub-Neptunes'' are larger $2.0$ - $3.5R_\oplus$ and typically have lower bulk densities, which suggest that they host significant primordial H/He atmospheres \citep[e.g.][]{Rogers2015,WolfgangLopez2015,JontofHutter2016}.}}

There are two ways to explain this contrast in atmospheric composition. In the first scenario, some planets are formed intrinsically rocky i.e. born without a H/He atmosphere, whilst others are able to accrete an atmospheric mass fraction of a few percent \citep[e.g.][]{NeilRogers2020,LeeConnors2020,Rogers2021}. In the second scenario however, it is suggested that after the gas accretion phase, some planets may undergo atmospheric mass-loss, such that they transition from a sub-Neptune to a super-Earth. In this scenario, lower-mass, more highly irradiated planets are more vulnerable to atmospheric loss. In essence, these two scenarios state that the radius gap was formed either at formation, or through evolution. {\jr{Alternatively, additional theories have been proposed to explain the radius gap. For instance, \citet{Zeng2019} argue that the sub-Neptunes could instead be dominated by water-ice, whereas \citet{Wyatt2020} discuss how atmospheric loss via impacts could contribute to the formation of the radius valley.}} Whilst in reality a combination of these scenarios is likely to occur, uncertainty still remains as to which mechanism dominates. Further, if the exoplanet population is strongly sculpted by mass-loss, it is debated which mechanism drives the planets to lose their primordial atmosphere. 

Of the possible atmospheric mass-loss models suggested, the two most prevalent are EUV/X-ray photoevaporation \citep[e.g.][]{Lammer2003,Baraffe2004,MurrayClay2009,Owen2013,Jin2014,LopezFortney2013,ChenRogers2016} and core-powered mass-loss \citep[e.g.][]{Ginzburg2016,Ginzburg2018,Gupta2019,Gupta2020}. In the photoevaporation model, the energy source for the mass-loss comes from the host star. Specifically the high energy EUV and X-ray flux from the star is capable of heating the upper atmospheres to high temperatures $\sim 10^4$~K, inducing a hydrodynamic outflow. Core-powered mass-loss, in contrast, is driven by a combination of stellar bolometric luminosity and remnant thermal energy from formation.
The latter results from a planet's accretion phase, during which gravitational binding energy is converted into heat. This thermal energy is slowly released from the core to the optically thick atmosphere and then radiated away. This results in a similar, but cooler hydrodynamic outflow. 

Despite the difference of energy source for each mechanism, the two models predict very similar observable signatures in the exoplanet demographics. As shown in multiple works, both models predict that the core compositions of super-Earths and sub-Neptunes are Earth-like \citep[e.g.][]{Owen2017,JinMordasini2018,Wu2019,Gupta2019,Rogers2021}. They also predict similar slopes to the radius valley as a function of orbital period and incident bolometric flux. Additionally, inference analysis suggests that the core mass distribution should be peaked at a few Earth masses for both models \citep{Gupta2019,Rogers2021}; however, it should be noted that the core-mass function derived from these models are different. If one is to determine which mechanism (if indeed there is one), that drives the evolution of close-in exoplanets, we must focus on predictions that differ between the models. One possible avenue is the temporal evolution of the radius distribution of super-Earths and sub-Neptunes.  Whilst photoevaporation predicts 
that the majority of mass-loss occurs in the first few $100$~Myrs after disc dispersal when the X-ray flux is stronger, core-powered mass-loss predicts this to occur on $\sim$~Gyr timescales. It follows that one would expect to observe stripped cores (i.e. super-Earths) around stars $\sim 100$~Myrs old if photoevaporation is taking place, and vice versa for core-powered mass-loss \citep{Gupta2020,Rogers2021}. Recent attempts to constrain the evolution of the radius gap have found that results are currently consistent with both models, in part because there is a significant lack of planets observed around young stars \citep{Berger2020, Sandoval2020}. With surveys such as THYME \citep{THYME-I,THYME-II,THYME-III,THYME-IV} additional young planetary systems will be revealed, allowing these predictions to be investigated further. 

In this paper, we present a new method for model comparison, which utilises a key difference between the two models. The energy source for photoevaporation stems from the host star, specifically from UV/X-rays. Crucially, the time-integrated X-ray exposure of a planet at a fixed incident bolometric flux decreases with stellar mass \citep[e.g.][]{McDonald2019}. This relationship arises due a combination of negative trends in stellar saturation time {\jr{(i.e. the time at which stars' ratio of high-energy to bolometric output begins to decline)}} with stellar mass, as well as in the ratio of X-ray-to-bolometric luminosity also with stellar mass \citep{Ribas2005,Wright2011,Jackson2012, Tu2015, Lopez2018, McDonald2019}. As a result, photoevaporation predicts that lower mass stars will be able to strip larger planets at fixed incident flux. Core-powered mass-loss on the other hand does not predict a trend in stellar mass at constant incident flux, assuming planets have similar ages and atmospheric opacities, providing a useful difference in demographic predictions. Note that since transit surveys like {\it Kepler} and {\it TESS} typically search for planets over a fixed range in orbital period, care must be taken to work at constant incident flux, not fixed orbital period, or orbital period ranges for this analysis to work correctly.

In Section \ref{sec:Method} we formalise this new method of model comparison and provide an overview of the two mass-loss mechanisms. These models are then used to make predictions of what one would expect to observe in exoplanet surveys. In Section \ref{sec:GradientCalc}, we discuss a method of extracting such trends from data and present results from synthetic and real surveys in Section \ref{sec:Results}. Discussions are presented in Section \ref{sec:Discussion} and we conclude in Section~\ref{sec:conclusions}.

\section{Method} \label{sec:Method}

\subsection{The Radius Gap in 3D} \label{sec:alpha&beta}
\begin{figure*} 
	\includegraphics[width=2.0\columnwidth]{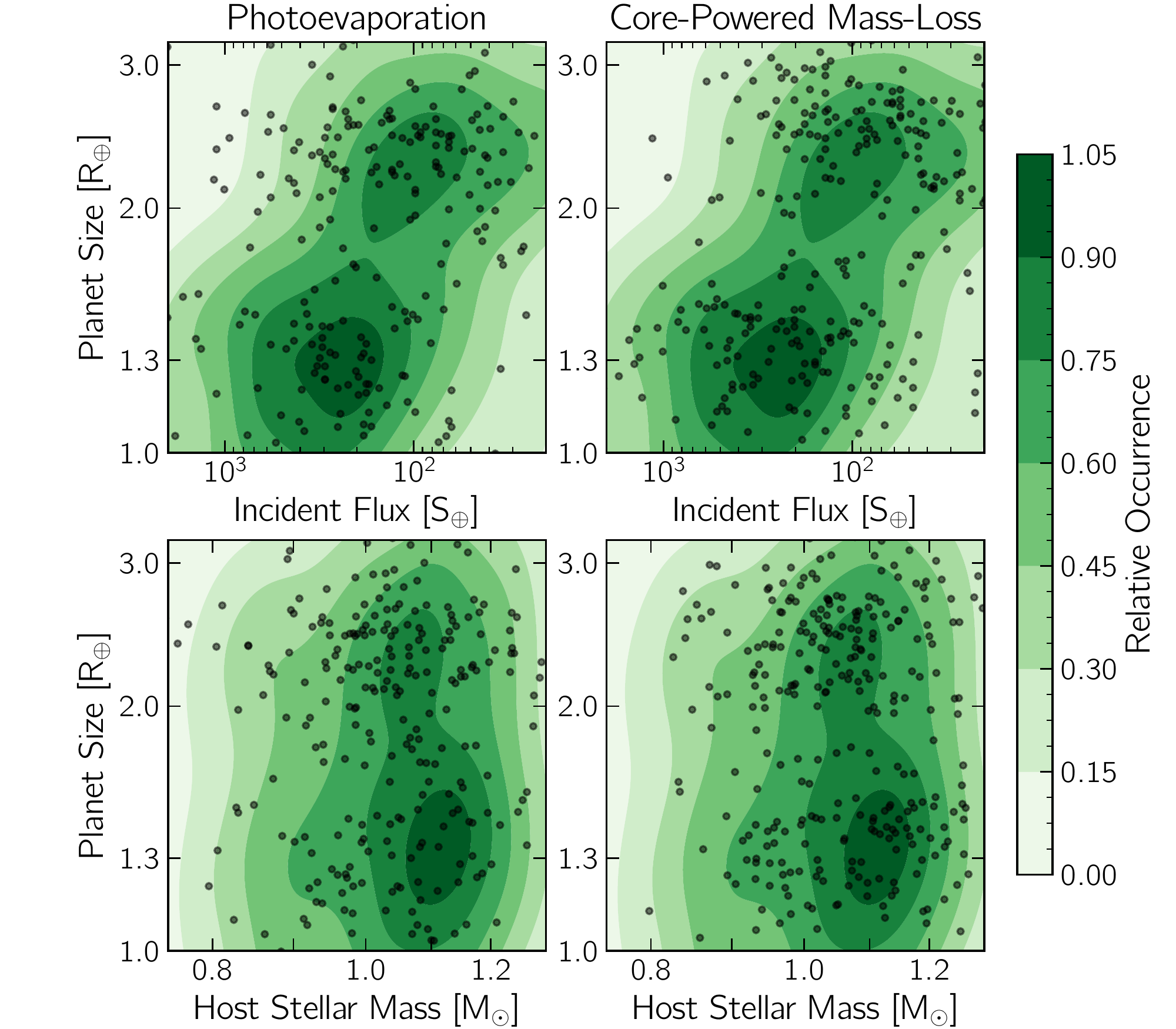}
    \caption{The radius gap is shown as function of incident bolometric flux (top row) and host stellar mass (bottom row). Here, green contours represent relative occurrence of the observed CKS planets \protect\citep[e.g.][]{Fulton2017}. Black points represent forward modelled planets from the photoevaporation model \protect\citep[e.g.][]{Owen2013} (left) and core-powered mass-loss model \protect\citep[e.g.][]{Ginzburg2016} (right). Note that modelled planets have appropriate bias and noise added such that they are representative of CKS data. Clearly, both models can reproduce the distribution of observed planets with the radius gap in the correct location and similar slopes. In order to determine which of the two models best describes the data, we must look to 3D representations of the radius gap, as laid out in Section \ref{sec:alpha&beta}. Details of prescriptions for each model are presented in Section \ref{sec:Models}.}
    \label{fig:SincPrad} 
\end{figure*}

As shown in Figure \ref{fig:SincPrad}, the models for photoevaporation and core-powered mass-loss are both capable of reproducing the general demographics of close-in exoplanets. Determining a method for model comparison therefore depends on finding observable signatures that are predicted to differ substantially from one model to another. In previous works, the radius gap has been quantified as a line in 2D parameter space (e.g. as a function of orbital period $P$, stellar mass $M_*$ or incident bolometric flux $S$, \citealt{Fulton2017,VanEylen2018,Berger2020}). However, both models predict the slopes in these planes to be very similar, and essentially indistinguishable with current and near-future exoplanet surveys. Crucially this is not the case when looking at the radius gap in higher dimensions, i.e. a plane as a function of stellar mass and incident bolometric flux. As discussed, this arises because the models disagree on how the radius gap behaves as a function of stellar mass at fixed incident bolometric flux. In order to quantify this, we define the position of the radius gap $R_\text{val}$ as the size of the largest super-Earths. This choice is made as this is the fundamental quantity that can be theoretically calculated from mass-loss models: namely, the largest core that can be entirely stripped at the position in parameter space\footnote{Although as we describe later this quantity cannot always be extracted from a noisy exoplanet population in an unbiased way.}.  Thus, for an arbitrary evolutionary model, we can describe the 3D radius gap as a joint power law of the form:
\begin{equation} \label{Eq:ModelJointPowerLaw}
    R_\text{val} \propto S^\alpha \; M_*^\beta,
\end{equation}
where $\alpha$ and $\beta$ are crucial mass-loss parameters and can be used to distinguish between photoevaporation and core-powered mass-loss. Specifically they are defined as:
\begin{equation} \label{Eq:AlphaBeta}
    \begin{split}
        \alpha & \equiv \left(\frac{\partial \log R_\text{val}}{\partial \log S}\right)_{M_*}, \\
        \beta & \equiv \left(\frac{\partial \log R_\text{val}}{\partial \log M_*}\right)_S .
    \end{split}
\end{equation}
From theoretical predictions (discussed in Section \ref{sec:ModelPredictions}) we expect $\alpha$ to be similar for both models, however we expect $\beta$ to be negative for photoevaporation and zero for core-powered mass-loss. This is because the high-energy exposure at fixed bolometric flux decreases as function of stellar mass \citep{McDonald2019}, which means that smaller stars will be able to strip larger cores at fixed incident bolometric flux in the photoevaporation model, thus implying a negative value of $\beta$. This crucial fact allows us to differentiate between the evolutionary histories, and use it as a tool to find signatures in the data. 

Before we use analytic models to predict theoretical values of $\alpha$ and $\beta$, we first address the radius gap in the $M_*$-$R_\text{p}$ plane. Specifically, we wish to show why using this plane to perform model comparisons is extremely difficult due to strong degeneracies in the models. We justify this claim with a simplistic approach. We start with the joint power law of Eq. \ref{Eq:ModelJointPowerLaw} in an inequality form, defining the maximum size of super-Earths for a given model, at that position in parameter space:
\begin{equation} 
    R_\text{val} \leq A \; S^\alpha \; M_*^\beta,
\end{equation}
where $A$ is some arbitrary constant. Next we aim to eliminate $S$ such that we can find the slope in the $M_*$-$R_\text{p}$ plane. To do so, we note that $S \propto L_* / a^2$ where $L_*$ is the bolometric luminosity, and utilise a stellar luminosity relation $(L_*/L_\odot) \propto (M_*/M_\odot)^\zeta$ as well as Kepler's third law $a^3 \propto P^2 M_*$ to find:
\begin{equation} 
    R_\text{val} \leq A' \; M_*^{\alpha (\zeta - \frac{2}{3}) + \beta} \; P^{-\frac{4}{3} \alpha},
\end{equation}
where $A'$ is a re-scaled constant. Noting that the period distribution of close-in exoplanets is approximately independent as a function of stellar mass \citep{Fressin2013,Dressing2015,LeeChiang2017,Petigura2018} and then differentiating we can derive the radius gap slope in the $P$-$R_\text{p}$ plane as:
\begin{equation} \label{Eq:RP_gradient}
    \frac{\text{d} \log R_\text{val}}{\text{d} \log P} \approx -\frac{4\alpha}{3}.
\end{equation}
Similarly, and more pertinent to this work, we can derive the radius gap slope in the $M_*$-$R_\text{p}$ plane as:
\begin{equation}\label{Eq:RM_gradient}
    \frac{\text{d} \log R_\text{val}}{\text{d} \log M_*} \approx \alpha (\zeta - \frac{2}{3}) + \beta.
\end{equation}
Inspection of this result shows that the slope is not only controlled by $\alpha$ and $\beta$ but also $\zeta$ and hence the stellar luminosity relation. Since $\zeta \gg |\beta|$ and $\zeta \gg \alpha$ for both models (as shown in Sections \ref{sec:PE-Predictions} and \ref{sec:CPML-Predictions}) the value of this slope is heavily controlled by the stellar luminosity relation \citep{Gupta2020,Loyd2020}.

\begin{figure*} 
	\includegraphics[width=2.0\columnwidth]{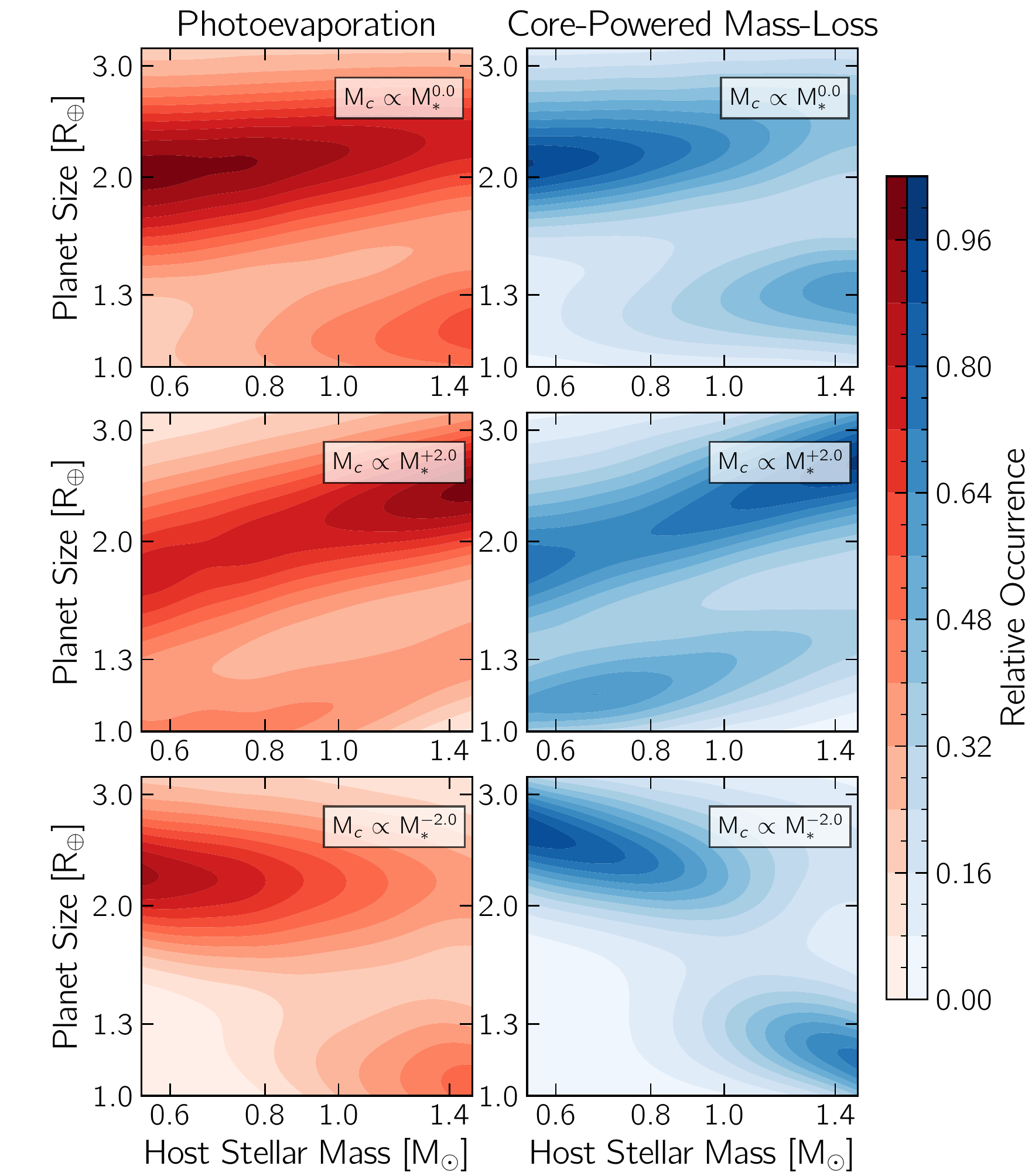}
    \caption{The radius gap is shown as a function of host stellar mass for the photoevaporation models (left) and core-powered mass-loss models (right). Here, contours represent relative occurrence and each slice in stellar mass is normalised to unity to better highlight the location of the radius gap. In the top row, we present the models for each of the two mass-loss schemes, where the core-mass function is independent of stellar mass. Both models predict a positive slope, however the exact value of slope is strongly controlled by the stellar mass-luminosity relation (see Section \ref{sec:alpha&beta}, specifically Eq. \ref{Eq:RM_gradient}). In the middle and bottom rows, we incorporate scalings of core mass to stellar mass into the models $M_\text{c} \propto M_*^{\pm 2}$, with positive and negative scalings in the middle and bottom row respectively. This clearly demonstrates that the radius gap slope can be fundamentally altered by the inclusion of such scalings. Although we do not argue scalings of these magnitudes are physical, we stress that attempting model comparisons and indeed demographic inferences in this parameter space are fraught with degeneracies.}
    \label{fig:SmassPrad} 
\end{figure*}

As an additional complication and as explored first by \cite{Wu2019}, we may invoke a core mass scaling with stellar mass which further contributes to the degeneracy in this result (see also \citealt{Gupta2020}). Although difficult to incorporate into Eq. \ref{Eq:RM_gradient}, the introduction of such scaling means that multiple combinations of $\alpha$, $\beta$, $\zeta$ and the unknown core-mass--stellar-mass scaling can yield the same value of the slope in the $M_*$-$R_\text{p}$ plane. To illuminate the effect of such a core-mass--stellar-mass scaling, Figure \ref{fig:SmassPrad} shows planet distributions of the photoevaporation and core-powered mass-loss models, calculated using the models presented in Section \ref{sec:Models}. In the top row, we show a set of planets for each model with a core-mass function that is independent of stellar mass, which clearly shows a positive slope in the $M_*$-$R_\text{p}$ plane. However, in the middle and bottom rows, we introduce core mass scalings $M_\text{c} \propto M_*^{\pm 2}$ with the positive scaling in the middle row and negative in the bottom row. Clearly, the slope of the radius-gap in this plane can be altered dramatically with the inclusion of such scalings, further contributing to the degeneracy in the slope. The reader should note however that the {\jr magnitudes of the scalings chosen/implemented here are not consistent with the current observations and} are used to simply demonstrate how correlations between stellar and planet masses can affect the slope.

Indeed the analysis of \cite{Wu2019}, in which demographic inferences were performed in the $M_*$-$R_\text{p}$ plane using the photoevaporation model, with the added introduction of a core mass scaling with stellar mass, argued for a linear relation between core-mass and stellar mass. However, we note that the implemented bolometric luminosity relation (i.e. $\zeta$ in Eq. \ref{Eq:RM_gradient}) was weaker than used in this work (and implemented in Figure~\ref{fig:MethodLimitation}).  {\jr We choose to adopt a value of $\zeta=4.5$ due to the fact that the CKS and GKS dataset are a magnitude-limited rather than a volume-limited sample, thereby including more high luminosity/mass stars \citep[][]{CKSI-Petigura2017,Gupta2020}. In addition, there exists a correlation between stellar mass and stellar metallicity in the \textit{Kepler} data \citep[e.g.][]{Eker2018,Owen2018,Petigura2018,Gupta2020} which, in combination with the latter, increases the value of $\zeta$ in the stellar luminosity relation from the standard empirical relation of $L_*\propto M_*^{3.2}$ \citep[e.g.][]{Cox2000} that was adopted in \cite{Wu2019}. Since a lower value of $\zeta$ yields a shallower radius gap slope in the $M_*$-$R_p$ plane (see Eq. \ref{Eq:RM_gradient}), the analysis of \cite{Wu2019} required a stronger core mass scaling in order to match the the observed value, compared to what one would infer from our implementation.} Thus, if this analysis were repeated with a larger value of $\zeta$, we speculate that the inferred stellar-to-core-mass scaling from the photoevaporation model would be weaker. Nevertheless, this type of inference will be useful in the future once constraints are placed as to which mechanism dominates atmospheric mass-loss, which we discuss in Section \ref{sec:Discussion}.

Thus, the clear benefit of analysing the radius gap in 3D is that in determining $\alpha$ and $\beta$, we do not integrate through $S$ or $M_*$ axes (as in the derivation for Eq. \ref{Eq:RM_gradient}) and hence avoid many of the degeneracies produced as a result. Additionally, the slopes $\alpha$ and $\beta$ are, to the first-order, independent of any stellar mass scalings. This is because the plane is defined as the maximum size of stripped cores for a given incident flux $S$ and stellar mass $M_*$. Hence, at constant $S$ and $M_*$, the only way that core mass to stellar mass scalings can manifest in the data is if there is an upper limit on the sizes of cores for a given stellar mass (i.e. higher mass stars can host planets up to a higher mass cutoff). If this were the case, super-Earths simply would not exist at certain stellar masses. Nonetheless, for scaling relations such as those considered in Figure \ref{fig:SmassPrad}, this 3D method provides a way of circumventing the effects of such scalings. We discuss this claim in Section \ref{sec:CoreMassScaling}.

\subsection{The Models} \label{sec:Models}

In this section, we first present analytic models for photoevaporation and core-powered mass-loss. We then use these to predict values of $\alpha$ and $\beta$ as we would expect to observe in demographic surveys.

Both analytic models describe how a planet that formed in a protoplanetary gas disc, evolves after disc-dispersal under either photoevaporation or core-powered mass-loss. Whilst the driver of photoevaporation comes from the high-energy flux from the host star, core-powered mass-loss uses a planet's primordial energy from accretion. This energy, in combination with the stellar bolometric luminosity eventually is able to drive the atmospheric outflow as the planet cools. In the subsequent sections, we provide a brief overview of both models, however for more information, we refer the reader to \citet{Owen2017} and \citet{Rogers2021} for photoevaporation and   \citet{Ginzburg2016,Ginzburg2018} and \citet{Gupta2019,Gupta2020} for core-powered mass-loss.

For both models, we assume that a planet of mass $M_\text{p}$ and radius $R_\text{p}$ has a core with Earth-like composition that is surrounded by a primordial H/He atmosphere. Following previous works \citep[e.g.][]{Piso2014,Lee2015,Inamdar2015} we assume that this atmosphere, which has a mass $M_\text{atm}$, has an inner convective layer which we assume to be adiabatic, and an outer radiative layer which is modelled as isothermal. The point where the atmosphere transitions between these regions is identified as the radiative-convective boundary $R_\text{rcb}$. In addition, as demonstrated by \citet[][]{Lopez2014} among other studies, we assume that most of the mass of such small exoplanets is in their cores (mass $M_\text{c}$, radius $R_\text{c}$) such that $M_\text{p} \sim M_\text{c}$. {\jr{It is inherently assumed that planets modelled under either mechanism initially go through a phase of rapid atmospheric mass-loss during protoplanetary disc dispersal, sometimes referred to as 'boil-off' or 'spontaneous mass-loss' \citep[e.g.][]{Ginzburg2016, Owen2016}. The reduction in atmospheric mass-fraction and size as a result of this process is incorporated into the initial conditions of both models.}}

Following previous observational studies \citep[e.g.][]{Fressin2013,Petigura2018}, both photoevaporation and core-powered mass-loss models assume that planets have the following underlying period distribution:
\begin{equation}{\label{eq:P_distr}}
\dv{N}{\;\text{log}P} \propto \begin{cases}
     P^{2}, & {P < 8\; \text{days}}, \text{ and} \\
    \text{constant}, & {P > 8\; \text{days}}.
  \end{cases}  
\end{equation}

\begin{figure} 
	\includegraphics[width=1.0\columnwidth]{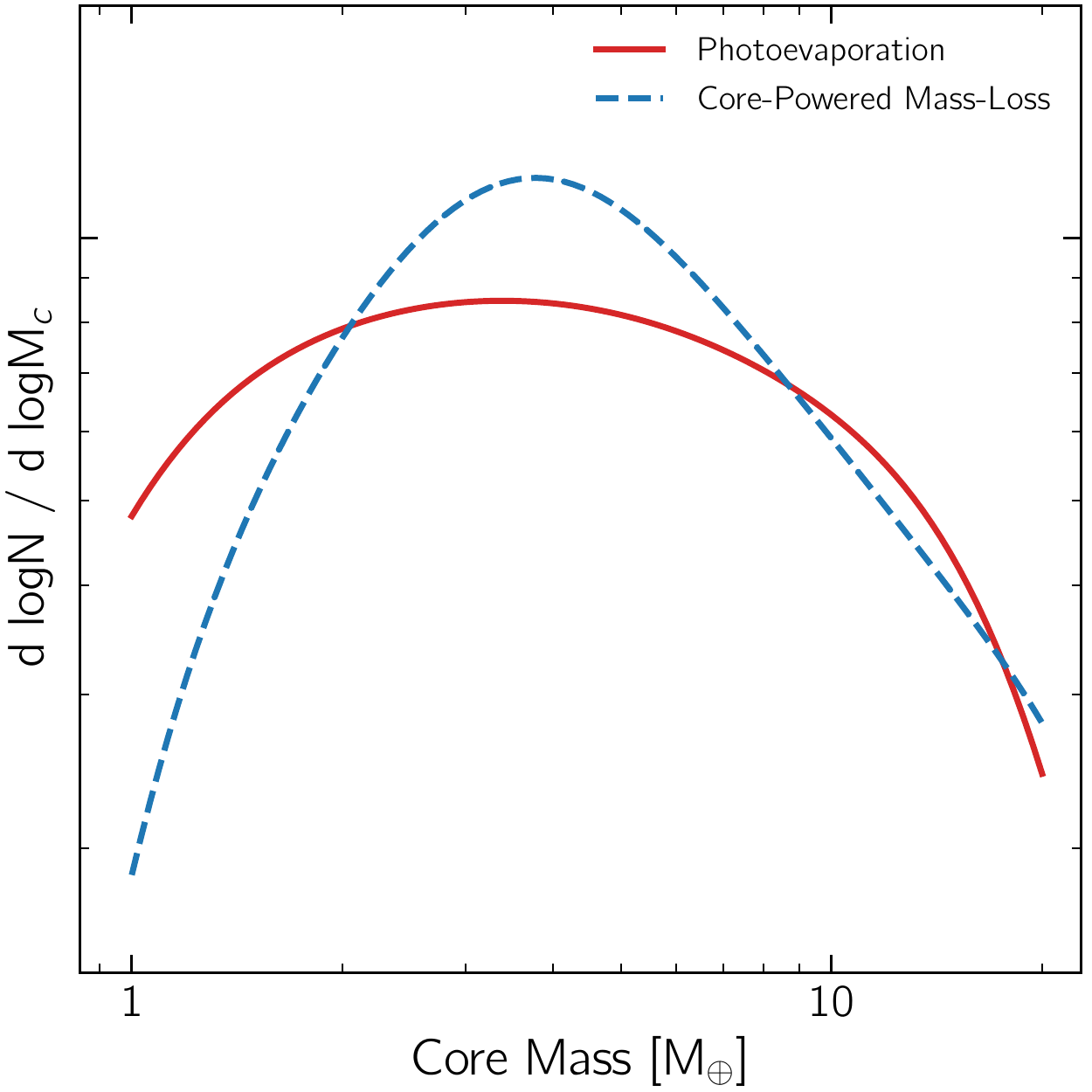}
    \caption{{\jr{The core mass distributions are shown for the photoevaporation model (solid line) and core-powered mass-loss model (dashed line). These are taken from the inference analysis of \protect{\citet{Gupta2019}} and \protect{\citet{Rogers2021}}, in which the respective models were fit to the CKS data \protect{\citep{Fulton2017}}.}}}
    \label{fig:CoreMass} 
\end{figure}

On the other hand, the models use different core mass distributions and initial atmospheric mass fraction distributions, which were determined through inference work focused on planets around Sun-like stars \citep{Gupta2020,Rogers2021}. {\jr{As a result, both models represent current best-fits to exoplanet observations. We show the core mass distributions of both models in Figure \ref{fig:CoreMass} and note that both distributions are consistent with current data. We emphasise that there would be little point performing our analysis with the same initial distributions as it would require at least one model to be inconsistent with the observations. Rather, this difference makes clear the importance of determining the dominant mass-loss model, as the inferred core-mass functions differ. }} {\jr{In other words, if the observational tests proposed in this paper reveal that  one mechanism dominates the planet evolution over the other mechanism, there will be little room to question such a conclusion as the models are already optimized to the observations.}} Further details of these distributions are discussed in the subsequent sections.

The evolutionary models for a planet's atmosphere are solved using the identical methods of \citet{Rogers2021} for the photoevaporation model and \citet{Gupta2019} for the core-powered mass-loss model, and we refer the curious reader to those works for the details. Essentially, the evolution models boil down to evolving the atmospheric mass fraction $X \equiv M_\text{atm} / M_\text{c}$ by numerically solving the following differential equation:
\begin{equation} \label{eq:ODE}
    \frac{\text{d}X}{\text{d}t} = -\frac{X}{t_{\dot{X}}},
\end{equation}
where $t_{\dot{X}}$ is the atmospheric mass-loss timescale, given by:
\begin{equation} \label{eq:PE-timescale}
    t_{\dot{X}} \equiv \frac{X}{\dot{X}} = \frac{M_\text{atm}}{\dot{M}_\text{atm}}.
\end{equation}
In the following sub-sections we detail how we determine the mass-loss rates in the different models. 

\subsubsection{EUV/X-ray Photoevaporation Model} \label{sec:PE}
In the photoevaporation mechanism, the mass-loss rate $\dot{M}_\text{atm}$ is calculated using the energy-limited mass-loss model \citep[e.g.][]{Baraffe2004,Erkaev2007}:
\begin{equation} \label{eq:massloss}
    \dot{M}_\text{atm} = \eta \; \frac{\pi R_\text{p}^3 L_\text{XUV}}{4 \pi a^2 GM_\text{c}},
\end{equation}
where $\eta$ is the mass-loss efficiency and parameterised with:
\begin{equation} \label{eq:masslossefficiency}
    \eta = \eta_0 \; \left( \frac{v_\text{esc}}{15\text{km s}^{-1}} \right ) ^{-\alpha_\eta}
\end{equation}
where $v_\text{esc}$ is the escape velocity, $\eta_0=0.1$ and $\alpha_\eta=2.0$ is the power-law index with numerical values taken from theoretical works of \cite{OwenJackson2012} and \cite{Owen2017}. We choose to take a different approach to stellar XUV luminosity evolution from previous works, as the accuracy of this has a large impact on the value of $\beta$ (shown in Section \ref{sec:PE-Predictions}). We assume that the ratio of XUV to bolometric luminosity follows a broken power law:
\begin{equation} \label{eq:LxLbol}
    \frac{L_\text{XUV}}{L_\text{bol}}=\begin{cases}
    \bigg( \frac{L_\text{XUV}}{L_\text{bol}} \bigg)_{\text{sat}} \; \bigg( \frac{M_*}{M_\odot} \bigg)^{-0.5} & \text{for } t < t_\text{sat}, \\
    \bigg( \frac{L_\text{XUV}}{L_\text{bol}} \bigg)_{\text{sat}} \; \bigg( \frac{M_*}{M_\odot} \bigg)^{-0.5} \; \bigg ( \frac{t}{t_\text{sat}} \bigg)^{-1-a_0}  & \text{for } t \geq t_\text{sat},
    \end{cases}
\end{equation}
where $a_0=0.5$ and $(L_\text{XUV} / L_\text{bol})_\text{sat} = 10^{-3.5}$ \citep{Wright2011, Jackson2012}. The saturation time $t_\text{sat}$ follows:
\begin{equation} \label{eq:tsat}
    t_\text{sat} = 10^{2} \; \bigg( \frac{M_*}{M_\odot} \bigg)^{-1.0} \;\; \text{Myr}.
\end{equation}
Note the departure from previous works \citep[e.g.][]{Owen2017,Rogers2021} with the introduction of negative stellar mass scalings in Eq. \ref{eq:LxLbol} and Eq. \ref{eq:tsat}, which is motivated by observational studies of \citet{Jackson2012,Shkolnik2014} and \citet{McDonald2019}. {\jr{We do not, however, incorporate the observed spread in $L_\text{XUV}/L_\text{bol}$ \citep[e.g.][]{Tu2015} as there is insufficient evidence as to how this varies as a function of stellar mass.}} Finally, in order to calculate $L_\text{XUV}$, we use \textsc{MIST} evolution tracks \citep{MIST-I2016, MIST-II2016} to accurately model the pre-main sequence and main-sequence evolution of the host star bolometric luminosity $L_\text{bol}$. As discussed in Section \ref{sec:TransitSurveys}, we eventually aim to simulate populations of planets that are akin to those in either the \textit{California-Kepler Survey} (CKS) \citep{CKSI-Petigura2017} or \textit{Gaia-Kepler Survey} (GKS) \citep{Berger2020}. As the CKS sample is a subset of the GKS sample with spectroscopic follow-up, we take the metallicities and masses of 100 randomly sampled host stars from this survey and model them through MIST evolution tracks. We then interpolate $L_\text{bol}$ evolution on a regular grid of time and stellar mass which is used in simulations of exoplanet photoevaporation. Not only does this improve the computational speed (instead of having to individually evolve each host star), but also means we naturally incorporate the intrinsic stellar luminosity relation into our simulations. {\jr{For completeness, we also simulated a set of planets through photoevaporation in which the stellar luminosity evolution was taken from \citet{Baraffe1998}. This yielded a difference in final planetary radii of $<0.2\%$, which is much less than typical measurement uncertainties. }}

The chosen core mass distribution and initial atmospheric mass fraction for photoevaporated planets comes from the inference analysis of \cite{Rogers2021}, in which $5^\text{th}$-order Bernstein polynomials were used to constrain their functional forms. The best fit distribution for core mass is peaked at $\sim 4M_\oplus$ with sharp drops in occurrence towards higher and lower mass cores. Similarly, the initial atmospheric mass fraction distribution is sharply peaked at $\sim 2\%$. Note that these two distributions are independent of each other i.e. initial atmospheric mass fraction is independent of core mass, which is not the case for the core-powered mass-loss model.

\subsubsection{Core-Powered Mass-Loss Model} \label{sec:CPML}

In the core-powered mass-loss mechanism, the mass-loss rate $\dot{M}_\text{atm}$ is given by: 
\begin{equation}
    \dot{M}_\text{atm}=\text{minimum}\bigg\{ \dot{M}_\text{atm}^E, \; \dot{M}_\text{atm}^B \bigg\}.
\end{equation}
where $\dot{M}_\text{atm}^E$ is the energy-limited rate i.e. the maximum mass-loss rate achievable assuming all cooling luminosity from the planet goes into driving atmospheric mass-loss,
\begin{equation}\label{eq:M_loss_rate_E}
\dot{M}_\text{atm}^E \simeq \frac{L_\text{rcb}(t)}{g R_c},
\end{equation}
where $L_\text{rcb}$ is the luminosity of the planet at the radiative-convective boundary, calculated assuming grey radiative diffusion. Alternatively, $\dot{M}_\text{atm}^B$ is the Bondi-limited rate i.e. the maximum mass-loss rate that is physically achievable given the thermal energy of the escaping gas molecules,
\begin{equation}\label{eq:M_loss_rate_B}
\dot{M}_\text{atm}^B = 4\pi R_\text{s}^2 c_\text{s} \rho_\text{rcb} \; \text{exp}\left( -\frac{GM
_\text{p}}{c_\text{s}^2 R_\text{rcb}}\right),
\end{equation}
where $R_\text{s}=GM_\text{c}/2c_\text{s}^2$ is the sonic radius for an isothermal sound speed $c_\text{s}$. Additionally, unlike in the photoevaporation model, we do not account for the pre-main sequence evolution of the host stars for core-powered mass-loss as this is only expected to have an observable impact on the planets evolving around host stars with $M_* \lesssim 0.5M_\odot$, which is below the range of stellar masses considered in this work.

Following \citet{Ginzburg2018} and \citet{Gupta2019}, we choose a core mass distribution that peaks at $\sim4 M_{\earth}$ and can be expressed as
\begin{equation} \label{eq:M_c_distr}
\dv{\;N}{\text{ log}M_\text{c}} \propto \begin{cases}
    M_\text{c}^2\;\text{exp} \left( -{M_\text{c}^2}/{(2 \sigma_{M_\text{c}}^2)} \right),&{M_\text{c} < 5\; M_\oplus}  \\
    M_\text{c}^{-1},&{M_\text{c} > 5\; M_\oplus},
  \end{cases}  
\end{equation}
where $\sigma_{M_\text{c}} \sim 3 M_\oplus$. Following \cite{Ginzburg2016}, in which gas accretion and subsequent mass-loss are modelled during disc dispersal, we assume that a planet's initial atmospheric mass fraction follows:
\begin{equation}\label{eq:f}
X \simeq 0.05 (M_\text{c}/M_\oplus)^{1/2}.
\end{equation}

\subsection{Model Predictions} \label{sec:ModelPredictions}

\subsubsection{Photoevaporation Predictions} \label{sec:PE-Predictions}
For photoevaporation, we derive $\alpha$ and $\beta$ in a similar manner to the work of \cite{Owen2017}, in which a scaling of the radius gap was determined as a function of orbital period and mean core density. In this case however, we aim to find a joint power-law of the form of Eq. \ref{Eq:ModelJointPowerLaw}. To begin, we state that for a given core mass, the longest evaporation timescale $t^{\text{max}}_{\dot{X}}$ is approximately equal to the stellar saturation time $t_\text{sat}$ i.e. the largest cores to be stripped are those which take the longest time, approximately equal to the saturation time of the host star. As argued by \citet{Owen2017} and \citet{Mordasini2020} planets that have not lost their atmosphere entirely by the end of the saturation period, retain them for the remainder of their lifetime. Combining definitions from Eq. \ref{eq:PE-timescale} with Eq. \ref{eq:massloss} and discarding order-unity constants, we may write the longest evaporation timescale as:
\begin{equation}
    t^\text{max}_{\dot{X}} \sim \frac{a^2 M_\text{p}^2 X_2}{\eta R_\text{}^3 L_\text{XUV}},
\end{equation}
where we set $X=X_2$, which is the atmospheric mass-fraction required to double a planet's core radius, and therefore yields the longest evaporation timescale \citep{Owen2017}. Equating this relation to $t_\text{sat}$ and rearranging we find a condition for the location of the radius gap:
\begin{equation}
    \frac{M_\text{val}^2 X_2}{R_\text{val}^3} \sim \eta t_\text{sat} \frac{L_\text{XUV}}{a^2}.
\end{equation}
We now incorporate all the necessary scaling of these variables with $R_\text{p}$, $S$ and $M_*$ such that we can find a joint power law of the form of Eq. \ref{Eq:ModelJointPowerLaw}. From its definition in Eq. 17 in \cite{Owen2017}, we write $X_2$ as:
\begin{equation}
    X_2 \propto T_\text{eq}^{-0.24} \, M_\text{c}^{0.17} \propto S^{-0.06} \, R_\text{p}^{0.68},
\end{equation}
where we have implemented a mass-radius relationship $M_\text{c} \propto R_\text{p}^4$ \citep{Valencia2006} for bare cores. We take the photoevaporation efficiency $\eta$ to scale with escape velocity, as in Eq. \ref{eq:masslossefficiency}, and hence:
\begin{equation}
    \eta \propto R_\text{c}^{-3}.
\end{equation}
Combining these relations with the scaling of $L_\text{XUV} / L_\text{bol}$ and $t_\text{sat}$ with stellar mass (Eq. \ref{eq:LxLbol} and \ref{eq:tsat} respectively), we find that:
\begin{equation}
    R_\text{val} \propto S^{\,0.12} \, M_*^{\,-0.17}.
\end{equation}
Finally, we find $\alpha$ and $\beta$:
\begin{equation} \label{Eq:AlphaBetaPE}
    \begin{split}
        \alpha & \equiv \left(\frac{\partial \log R_\text{val}}{\partial \log S}\right)_{M_*} \simeq  0.12, \\
        \beta & \equiv \left(\frac{\partial \log R_\text{val}}{\partial \log M_*}\right)_S \;\,\, \simeq  -0.17.
    \end{split}
\end{equation}
Putting these values into the prediction for the slope of the radius gap in the $M_*$-$R_\text{p}$ plane from Eq. \ref{Eq:RM_gradient}, we find a predicted value of:
\begin{equation}\label{Eq:RM_gradient_PE}
    \frac{\text{d} \log R_\text{val}}{\text{d} \log M_*} \approx 0.29,
\end{equation}
where we have assumed a stellar luminosity relation $(L_*/L_\odot) \propto (M_*/M_\odot)^\zeta$ with $\zeta = 4.5$ consistent with the CKS data set (\citealt{CKSI-Petigura2017,Gupta2020}, {\jr{see Section \ref{sec:alpha&beta})}}. Additionally, and as shown in \citet{Owen2017}, the value of the slope in the $P$-$R_\text{p}$ plane from Eq. \ref{Eq:RM_gradient_PE} is predicted to be:
\begin{equation}\label{Eq:RP_gradient_PE}
    \frac{\text{d} \log R_\text{val}}{\text{d} \log P} \approx -0.16,
\end{equation}

Note that the assumptions made in this photoevaporation model, particularly the scaling of $t_\text{sat}$ and $L_\text{XUV} / L_\text{bol}$ with stellar mass, as well as the efficiency scaling were chosen to the yield the most conservative estimate of $\beta$. This is because we wish to create a scenario in which differentiating between the two models is as challenging as reasonably possible. Changing these scalings to produce a more extreme value of $\beta$ may falsely suggest that the models differ more substantially. As the evolution of high energy stellar luminosity still remains uncertain, we choose to adopt values that produce a prediction of $\beta$ that is conservative, whilst still physically justifiable. We discuss these assumptions in Section \ref{sec:modelparams}.

\subsubsection{Core-Powered Mass-Loss Predictions} \label{sec:CPML-Predictions}
To determine a joint power-law of the form Eq. \ref{Eq:ModelJointPowerLaw} and hence derive $\alpha$ and $\beta$ for core-powered mass-loss, we appeal to the argument previously presented by \citet{Gupta2019} in which the slope of the radius valley in the planet distribution is set by the condition when the cooling timescale equals the mass-loss timescale, $t_\text{cool} = t_{\dot{X}}$. \citet{Gupta2019} showed that because $t_{\dot{X}}$ has an exponential dependence whereas $t_\text{cool}$ doesn't, this equality can be approximated to $G M_\text{c}/c^2_\text{s} R_\text{rcb} = \text{constant}$. If we then substitute for $c_\text{s}$, use the mass-radius relation and approximate $R_\text{rcb} \simeq R_\text{p} \sim 2 R_\text{c}$, this relation reduces to $R_\text{p}^3 T_\text{eq}^{-1} \simeq \text{constant}$. Changing variables from equilibrium temperature to incident bolometric flux yields:
\begin{equation} \label{eq:slope_CPML}
   R_\text{val}^3 T_\text{eq}^{-1} \sim R_\text{val}^3 S^{-1/4} \simeq \text{constant}.
\end{equation}
Hence the predictions of $\alpha$ and $\beta$ for core-powered mass-loss are as follows:
\begin{equation} \label{Eq:AlphaBetaCPML}
    \begin{split}
        \alpha & \equiv \left(\frac{\partial \log R_\text{val}}{\partial \log S}\right)_{M_*} \simeq  0.08, \\
        \beta & \equiv \left(\frac{\partial \log R_\text{val}}{\partial \log M_*}\right)_S \;\,\, \simeq  0.00,
    \end{split}
\end{equation}
Note that due to lack of a stellar mass term in Eq. \ref{eq:slope_CPML}, the predicted value of $\beta$ is zero. As with photoevaporation in Eq. \ref{Eq:RM_gradient_PE}, we may use these values to predict the slopes of the radius gap in the $M_*$-$R_\text{p}$ and $P$-$R_\text{p}$ planes from Eqs. \ref{Eq:RM_gradient} and \ref{Eq:RP_gradient}, yielding:
\begin{equation}\label{Eq:RM_gradient_CPML}
    \begin{split}
    \frac{\text{d} \log R_\text{val}}{\text{d} \log M_*} & \approx 0.32, \\
    \frac{\text{d} \log R_\text{val}}{\text{d} \log P} & \approx -0.11,
    \end{split}
\end{equation}
which are consistent with the analytical and numerical results of \citet{Gupta2020}, with the slope in $M_*$-$R_\text{p}$ plane arising simply as a by-product of the stellar mass-luminosity correlation. Comparing these predictions to values of 0.29 and -0.16 for photoevaporation (Eqs. \ref{Eq:RM_gradient_PE} and \ref{Eq:RP_gradient_PE}) further demonstrates that performing model comparisons in either of these 2D parameter spaces is extremely challenging.

\subsection{Synthetic Transit Surveys} \label{sec:TransitSurveys}
Before attempting to measure $\alpha$ and $\beta$ from the real data, we first seek to determine whether they can be measured from synthetic surveys created using either the photoevaporation or core-powered mass-loss models. Evidently, the larger the survey i.e. the more planets that are observed, the better the constraints on $\alpha$ and $\beta$ are likely to be. However, unlike with analysis of the radius gap in 2D \citep[e.g.][]{Fulton2017,VanEylen2018,Berger2020}, more planets are needed to analyse the gap in 3D due to geometric scaling of the statistical problem. Additionally, effects such as the stellar mass distribution, measurement uncertainty and false positive contamination will also have an effect on the extraction of $\alpha$ and $\beta$.

In order to understand all these effects, we produce synthetic transit surveys which model the completeness and noise of such an observation. By controlling the number of planets that are observed, we therefore also control the Poisson noise. To do this, we model 100,000 planets and follow a similar prescription to \cite{Rogers2021}, in which a high accuracy PDF for planet occurrence is calculated from each model in $S$-$M_*$-$R_\text{p}$ space using a Kernel Density Estimation. We create the PDF using the kernel density estimator \textit{fastKDE} of \cite{Obrien2014,Obrien2016} as this is computationally faster than standard techniques such as using a Gaussian kernel. We adopt an underlying stellar mass distribution as a Gaussian of mean $\mu_{M_*}$ and standard deviation $\sigma_{M_*}$. We investigate the effect of changing this distribution in Section \ref{sec:Results}. The PDF for planet occurrence is then multiplied by the completeness map for the \textit{Kepler} survey taken from \citet{Fulton2017}, but now extended to 3D space to match the domain of the planet occurrence PDF (see \citet{Rogers2021} for details on this calculation). Note that in doing so, we have assumed completeness is approximately constant as a function of stellar mass in the range $0.5M_\odot < M_*< 1.5M_\odot$. This multiplication has the effect of biasing the synthetic survey to preferentially observe planets with large radii and high incident bolometric flux. The new PDF describes a high accuracy representation of planet detection for a \textit{Kepler}-like transit survey. The final task is to draw a desired number of planets $N_\text{p}$ from this 3D PDF and incorporate measurement uncertainty. These are incorporated by adding a random Gaussian perturbation to planet parameters with zero mean and standard deviation equal to the associated error. Finally, we take these synthetically observed planets and measure $\alpha$ and $\beta$ as laid out in Section \ref{sec:GradientCalc}. In Section \ref{sec:Results}, we aim to quantify how measurement uncertainty, false positive contamination and Poisson noise effect results of $\alpha$ and $\beta$. Hence to quantify this, we repeat the above process 10,000 times to produce joint-posterior distributions for $\alpha$ and $\beta$ for a given survey setup.

\subsection{Gradient Calculation} \label{sec:GradientCalc}

\begin{figure*} 
	\includegraphics[width=2.0\columnwidth]{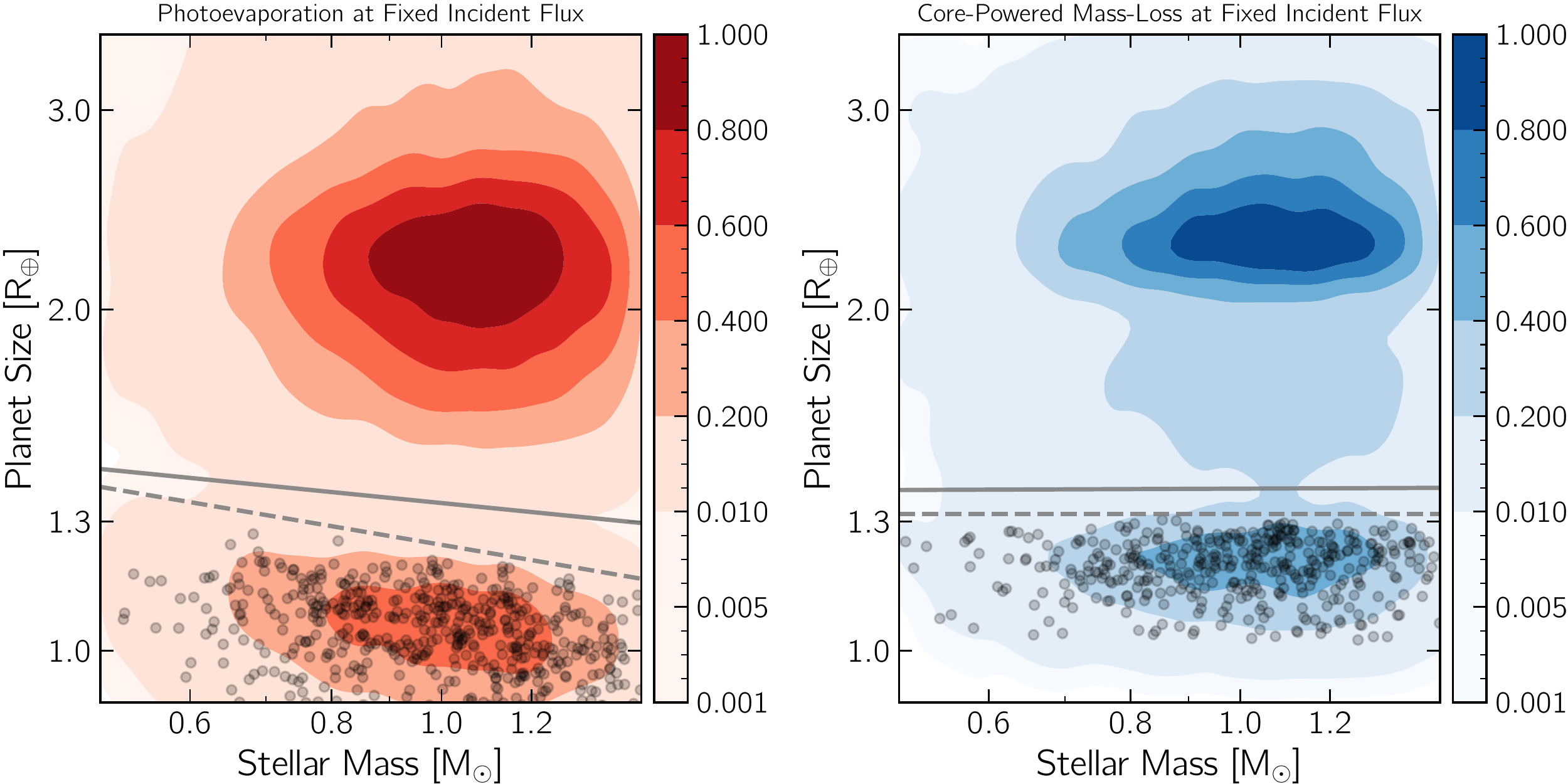}
    \caption{The underlying planet distributions for photoevaporation (left) and core-powered mass-loss (right) in the $M_*$- $R_\text{p}$ plane at constant incident flux $S=50S_\oplus$. By holding incident flux constant, the radius gap slope is representative of $\beta$. Here contours show relative occurrence. Dashed lines show theoretical predictions of $\beta$ from Sections \protect{\ref{sec:PE-Predictions}} and \protect{\ref{sec:CPML-Predictions}}, whilst solid lines show values of $\beta$ measured using method described in Section \protect{\ref{sec:GradientCalc}}. These two lines are offset because the theoretical radius gap defines the size of the largest super-Earths and hence is at lower radii than gap measured by our algorithm. Individual super-Earths are plotted for each model in black i.e. those with no atmosphere present after 3~Gyrs of evolution, in the incident flux range $45\;S_\oplus < S < 55\;S_\oplus$.}
    \label{fig:MethodLimitation} 
\end{figure*}

Determining the position and slopes of a radius gap in real data is an ill-defined statistical problem. Whereas we typically wish to find where data trends exist, we now aim to quantify a trend in the absence of data. Here we present a methodology for isolating and extracting $\alpha$ and $\beta$ from data in the 3D space of $S$-$M_*$-$R_\text{p}$. Instead of a line, we now characterise the valley as a plane in log-space in the following manner:
\begin{equation}
    \begin{split}
        \log R_\text{val} & = \left(\frac{\partial \log R_\text{val}}{\partial \log S}\right)_{M_*} \log S \,
         + \, \left(\frac{\partial \log R_\text{val}}{\partial \log M_*}\right)_S \log M_* \,+ \,\log R_0,
    \end{split}
\end{equation}
where $\log R_0$ is the normalisation of the plane and a nuisance parameter in this problem. We can then use definitions of $\alpha$ and $\beta$ from Eq. \ref{Eq:AlphaBeta} to simplify: 
\begin{equation} \label{Eq:plane}
    \log R_\text{val} = \alpha \log S + \beta \log M_* + \log R_0.
\end{equation}
As we expect the radius valley to be flat in log-space, we expect $\alpha$ and $\beta$ to be constants and can therefore extract them to quantify the slope for a given set of planets. The methodology for this fitting is as follows: we first use a population of observed planets (such as those produced from the synthetic surveys in Section \ref{sec:TransitSurveys}) to calculate a PDF for planet detection in $S$-$M_*$-$R_\text{p}$ space. As with Section \ref{sec:TransitSurveys}, we create the PDF using \textit{fastKDE} of \cite{Obrien2014,Obrien2016}. We label this PDF as $\lambda(S,M_*,R_\text{p})$. We then take slices at constant stellar mass $M_{*,i}$ through this PDF and locate the radius gap in the $S$-$R_\text{p}$ plane for each 2D slice. To do this, we define the radius gap at slice $i$ as:
\begin{equation} \label{Eq:line}
    \log R_\text{val} = \alpha_i (\log S-2) + \log R_{\text{p,}100,i}
\end{equation}
where $\log R_{\text{p,}100,i}$ is the intercept at $S=100S_\oplus$. We extract parameters $\alpha_i$ and $\log R_{\text{p,}100,i}$ for each slice by evaluating the line integral $\mathcal{I}$ through $\lambda(S,M_*,R_\text{p})$:
\begin{equation} \label{Eq:surfaceInt}
    \mathcal{I} = \int_\mathcal{C} \lambda(S,M_{*,i},R_\text{p}) \; \text{d}S,
\end{equation}
where $\mathcal{C}$ is the path defined by the line in Eq. \ref{Eq:line}. The exact parameters are determined by minimising $\mathcal{I}$ using \textit{scipy.optimize}, which provide $\alpha_i$ and $\log R_{\text{p,}100,i}$ for each slice. The final value of $\alpha$ is calculated as the mean $\alpha_i$ for all slices, whilst the value for $\beta$ is determined by fitting a straight line through the values of $\log R_{\text{p,}100,i}$ as a function of stellar mass $M_{*,i}$. The slope of this line provides a value for $\beta$. In the event that this method misidentifies the radius gap for a given slice, the value of $\alpha_i$ or $\log R_{\text{p,}100,i}$ will be an outlier compared to those of other slices in the same PDF. We employ a simple outlier detection scheme to remove these erroneous values, by removing any $\alpha_i$ or $\log R_{\text{p,}100,i}$ that lies $2\sigma$ away from the mean value, where $\sigma$ is the standard deviation of $\{\alpha_i\}$ or $\{\log R_{\text{p,}100,i}\}$. In this work we choose $N=15$, with slices logarithmically spaced between $1-\sigma_{M_*}$ and $1+\sigma_{M_*}$ where $\sigma_{M_*}$ is the standard deviation of the stellar mass distribution. It was determined that extracted values of $\alpha$ and $\beta$ converged with $N \gtrsim 10$, which is reflected in our choice of $N$.

Although in this work we apply this method to the 3D space of $S$-$M_*$-$R_\text{p}$, it can be used to locate the radius gap in any chosen parameter space. To demonstrate its effectiveness and also its limitations, Figure \ref{fig:MethodLimitation} shows the gradients calculated in the 2D $M_*$-$R_\text{p}$ plane at a constant incident flux for photoevaporation and core-powered mass-loss models. Note that by holding incident flux constant the radius gap slope is representative of $\beta$, not that of Eq. \ref{Eq:RM_gradient} and those shown in Figure \ref{fig:SmassPrad}. Solid lines show the slopes extracted by minimising the PDF integral through the $M_*$-$R_\text{p}$ plane, whilst dashed lines represent the theoretical predictions of $\beta$ for the respective models. This highlights a limitation of the methodology, which is particularly visible for the photoevaporation data. Note that whilst the predicted and measured slopes are equal for core-powered mass-loss data, they differ for photoevaporation, with the measured value being shallower than theoretically predicted. This issue arises because the numerical method for determining the slope locates the line that bisects the two populations of super-Earths and sub-Neptunes, whilst the theoretical predictions are determined by finding the size of the largest super-Earths. This subtle distinction implies that the we expect the extracted value of $\beta$ to be shallower\footnote{Note that if the line of smallest sub-Neptunes is parallel to the line of largest super-Earths then the measured value will be approximately equal to that which is theoretically predicted. This is the case for $\alpha$ and $\beta$ in the core-powered mass-loss model but only $\alpha$ for photoevaporation.} than the theoretical prediction of $\beta=-0.17$. The implications of this effect are discussed in Section \ref{Sec:Limitations}. We emphasise that in this work we present a {\it possible} method for extracting the values of $\alpha$ and $\beta$ from an observed population of exoplanets that successfully works on real data, albeit in a biased fashion. Therefore, while the theoretical ideas underpinning the observational test are independent of the method used, the accuracy and precision to which these values can be extracted from the same data-set is method dependent. Since our best method is biased, we encourage further work to develop a precise and unbiased method.

\section{Results} \label{sec:Results}

\begin{figure*} 
	\includegraphics[width=1.8\columnwidth]{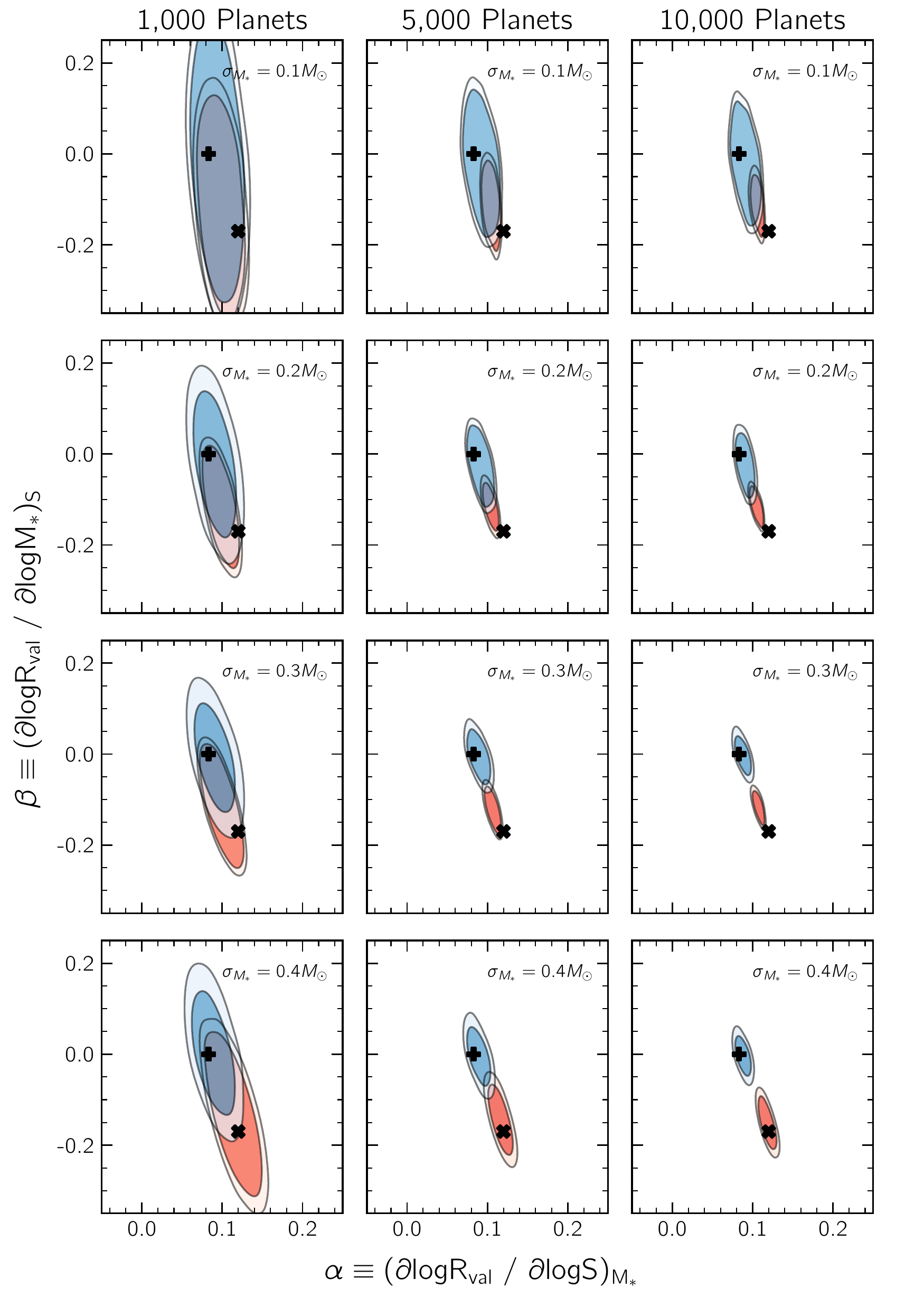}
    \caption{Posterior distributions are shown of key radius gap parameters $\alpha$ and $\beta$ for synthetic transit surveys with planets modelled through photoevaporation evolution in red and core-powered mass-loss evolution in blue. Contours represent the $1\sigma$ and $2\sigma$ levels, with expected values from theoretical predictions shown as a `$+$' for core-powered mass-loss and a `$\times$' for photoevaporation. We demonstrate that $\alpha$ and $\beta$ are better constrained for larger surveys (moving from left to right) and wider stellar mass distribution widths (moving from top to bottom). Note that the posteriors are not always consistent with theoretical predictions due to limitations in the gradient extraction method (see Sections \ref{sec:GradientCalc} and \ref{Sec:Limitations}).}
    \label{fig:Summary} 
\end{figure*}

\begin{figure} 
	\includegraphics[width=1.0\columnwidth]{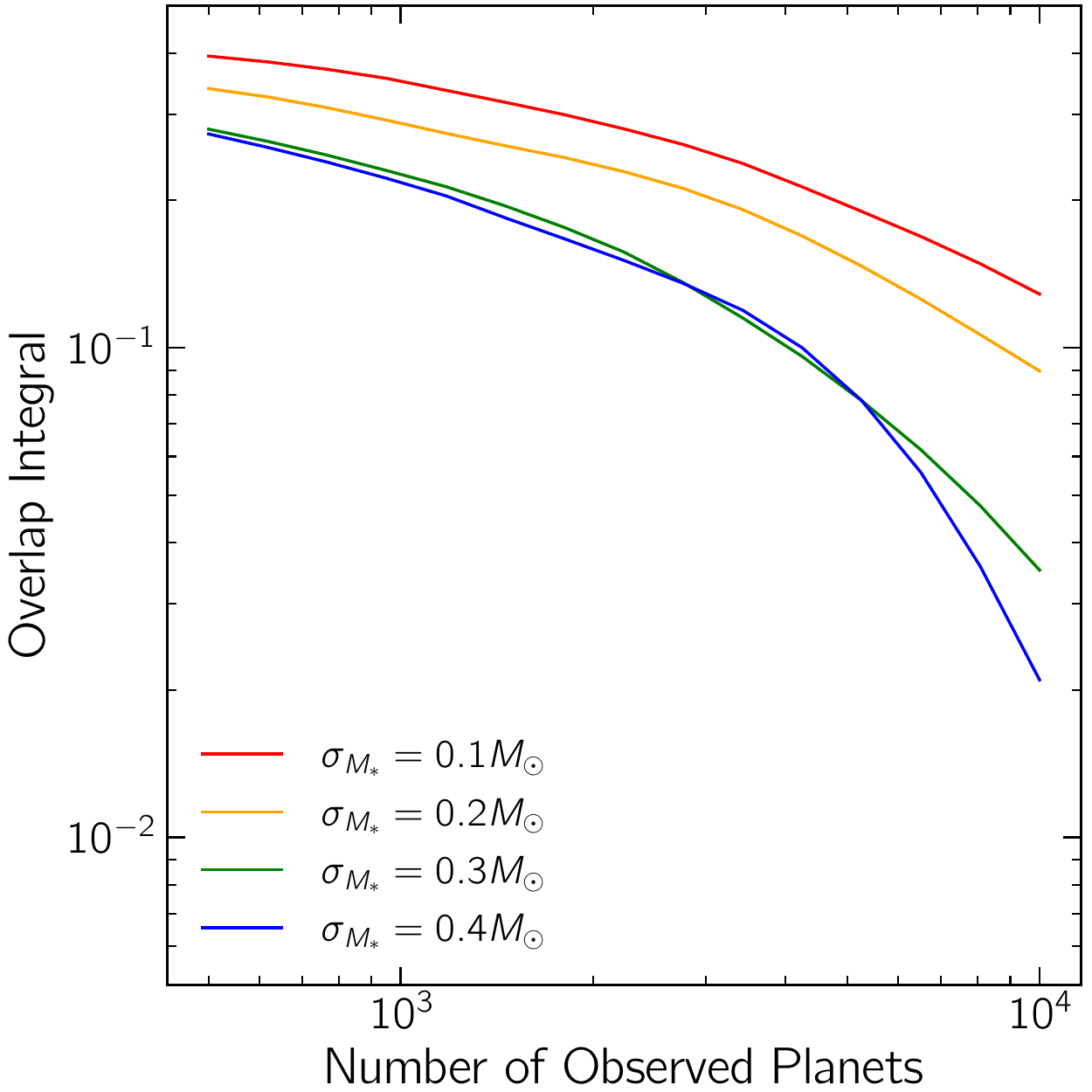}
    \caption{The overall limitations of extracting radius gap parameters $\alpha$ and $\beta$ for synthetic transit surveys are demonstrated with the overlap integral between posteriors of Figure \protect{\ref{fig:Summary}} as function of survey size. This captures the ``precision" of the statistical analysis i.e. how much do the posteriors overlap and prevent separation of models. Essentially this is a measure of the probability that one would fail to distinguish between the models in a given experiment. Different coloured lines show the overlap integral for varying widths in stellar mass distribution.}
    \label{fig:OverlapNominal} 
\end{figure}

The goal of this work is to determine the qualities of a hypothetical transit survey suitable for determining which model best describes the exoplanet data. Therefore, we wish to ascertain approximate values for the required number of detected planets $<4R_\oplus$, measurement uncertainty threshold, false positive contamination level and also the host stellar mass distribution for such a survey. To begin, we reproduce a CKS \citep{Fulton2018} or GKS \citep{Berger2020} style survey which both benefit from stellar radii constraints from \textit{Gaia} \citep{Gaia2018}. As a result, we model the typical measurement uncertainties of 5\% for planet radius, 5\% for stellar mass and 8\% for incident bolometric flux. We perform surveys with a host stellar mass distribution approximated as a Gaussian function with mean $\mu_{M_*} = 1.0M_\odot$ and varying standard deviation $\sigma_{M_*} = \{0.1, 0.2, 0.3, 0.4 \} M_\odot$. Initially we also assume that there are no false positives in the survey. Recall that the two parameters of interest are $\alpha = (\partial \log R_\text{val} / \partial \log S)_{M_*}$ and $\beta = (\partial \log R_\text{val} / \partial \log M_*)_{S}$. As $\beta$ is the parameter expected to differ the most between the two models and quantifies how the radius gap behaves as a function of stellar mass, it follows that the wider the stellar mass distribution, the better the constraint. Recall that for a survey of size $N_\text{p}$ planets, we quantify the uncertainty in these parameters by redrawing $N_\text{p}$ planets and extracting $\alpha$ and $\beta$ 10,000 times. This is performed twice: once for planets modelled using the photoevaporation scheme and again for planets modelled using the core-powered mass-loss scheme.

Figure \ref{fig:Summary} shows a summary of posteriors of $\alpha$ and $\beta$ for varying size of survey and host stellar mass distribution width. Red contours represent the $1\sigma$ and $2\sigma$ bounds for synthetic surveys with photoevaporated planets, whilst blue contours show the same for planets that have undergone core-powered mass-loss. Black crosses represent the theoretical values expected from the each model, given by Eq. \ref{Eq:AlphaBetaPE} for photoevaporation and Eq. \ref{Eq:AlphaBetaCPML} for core-powered mass-loss. As expected, the size of the posteriors reduces as the size of the survey increases (i.e. moving right in the figure). This can be understood because the increase in detected planets results in the Poisson noise having less of an effect on the planet distribution. Additionally, as the width of the stellar mass distribution is increased (i.e. moving down in the figure), the contours move closer to the theoretical predictions of $\alpha$ and $\beta$. Again, this is expected, as a wider distribution provides more information on how the radius gap changes as function of stellar mass. Note how the posteriors for photoevaporation are consistently finding a shallower value for $\beta$ than theoretically predicted, as discussed in Section \ref{sec:GradientCalc}, due to the inherent bias in extracting these slopes; see Section \ref{Sec:Limitations} for further discussion. 

To quantify the effectiveness of the proposed statistical method of extracting $\alpha$ and $\beta$, we determine the overlap integral of the posteriors shown in Figure \ref{fig:Summary}. The overlap integral assesses the ``precision'' of the method i.e. how well can photoevaporation be distinguished from core-powered mass-loss. This is shown in Figure \ref{fig:OverlapNominal} as a function of survey size and stellar mass distribution width. It can be seen that the overlap decreases with increasing size of survey as well as increasing stellar mass distribution width, confirming the general trends seen in Figure \ref{fig:Summary}. Additionally, the overlap integral estimates the significance level at which the two posteriors can be distinguished. As an example, for the survey with stellar mass width $\sigma_{M_*} = 0.3M_\odot$, we can differentiate between the two models $90\%$ of the time with a survey of $\gtrsim 5000$ planets.

\subsection{Measurement Uncertainty}
An important question is at what threshold of measurement uncertainty can $\alpha$ and $\beta$ to be sufficiently extracted from a given survey. In order to do this, we take a baseline survey with a stellar mass distribution of a Gaussian function, centred at $1M_\odot$ with a standard deviation of $0.3M_\odot$ and then vary the fractional measurement uncertainty to see how this effects the analysis. We consider three scenarios: the first in which percentage errors for each of our quantities $(S, M_*, R_\text{p})$ are $(4\%, 2\%, 2\%)$, a second scenario with $(8\%, 5\%, 5\%)$ and finally a third with $(12\%, 8\%, 8\%)$. Note that uncertainties in incident flux $S$ from transit surveys are typically larger as this quantity is determined via stellar modelling which is used to calculate the bolometric luminosity and hence incident flux.

\begin{figure} 
	\includegraphics[width=1.0\columnwidth]{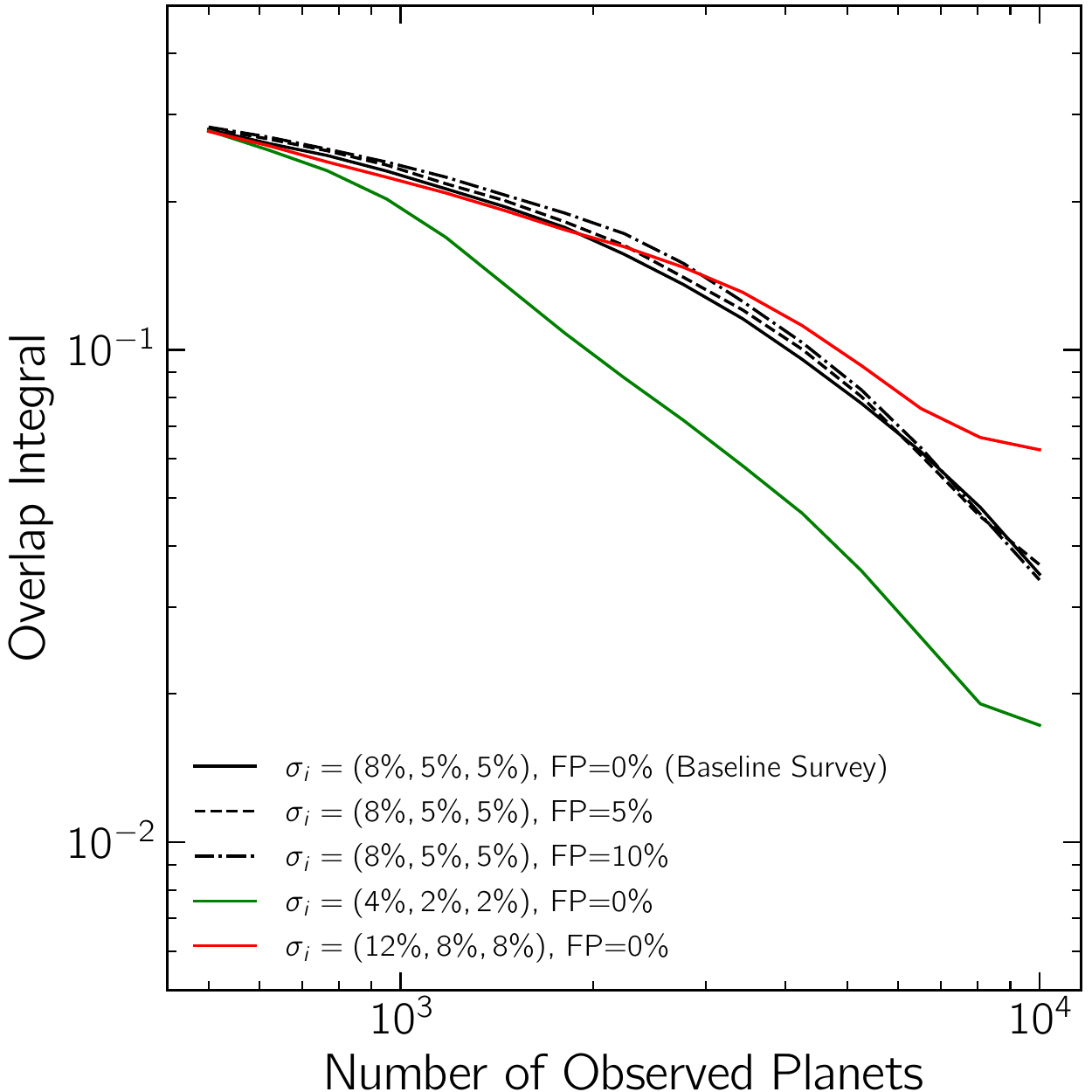}
    \caption{Same as Figure \protect{\ref{fig:OverlapNominal}}, whereas here the overlap integral is calculated for varying survey quality as a function of survey size. Here, $\sigma_i$ refers to the percentage errors in each of three measurements required for this statistical analysis $(S,M_*,R_\text{p})$, whilst FP represents the percentage of false positives present in the survey. The baseline survey (black solid line) has a stellar mass distribution standard deviation of $0.3M_\odot$ (same as green line in Figure \protect{\ref{fig:OverlapNominal}}). Dashed and dot-dashed lines represent the baseline survey with a 5\% and 10\% false positive contamination. Red and green lines represent the baseline survey with measurement uncertainties of $(4\%, 2\%, 2\%)$ and $(12\%, 8\%, 8\%)$ respectively.}
    \label{fig:OverlapUncFP} 
\end{figure}

Figure \ref{fig:OverlapUncFP} demonstrates the effect of changing the measurement uncertainties of a survey by considering the overlap integral. In green, it shows that reducing percentage uncertainties to $(4\%, 2\%, 2\%)$ can dramatically improve the distinction between photoevaporation and core-powered mass-loss models by up to $\sim50\%$. Increasing the uncertainties on the other hand, to $(12\%, 8\%, 8\%)$, has the opposite effect with a larger overlap between the two posteriors shown in red. The result can be understood because measurement uncertainty can stochastically shift planets into the radius gap and thus mask the underlying values of $\alpha$ and $\beta$.

\subsection{False Positives} \label{sec:FP}
Similar to measurement uncertainty, the presence of false positives in a sample can place data points into the radius gap, and thus reduce a survey's purity. To investigate this effect, we model false positive contamination as data points randomly drawn within $S\in[10^{-1},10^{4}]S_\oplus$, $M_*\in[0.5,1.5]M_\odot$ and $R_\text{p} \in [1.0,4.0]R_\oplus$. For a given survey of size $N_\text{p}$, we remove a fraction of these planets and replace them with false positives. We show the results for a 5\% and 10\% contamination in dashed and dot-dashed lines respectively in Figure \ref{fig:OverlapUncFP}. Interestingly, the injection of false positives have little, to no effect on the the ability to extract $\alpha$ and $\beta$ and thus on the distinction between photoevaporation and core-powered mass-loss. We can understand this as follows, whereas measurement uncertainty effects every data point in the survey and crucially those near the radius gap, false positives are injected randomly across the whole domain, meaning a large number of false positive will be placed far from the gap. Thus, up to at least a 10\% contamination level, any false positives found in the gap will not prevent $\alpha$ and $\beta$ from being extracted accurately.

\subsection{Changing Stellar Mass Distribution}
So far in this study, we have restricted the stellar mass distribution to be centred at $1M_\odot$, as this is typical of the CKS or GKS datasets \citep{Fulton2017, Berger2020}. Further improvements can be made in our ability to distinguish between the two models if the stellar mass distribution shifts to lower masses. To understand this, we must take a closer look at the radius gap. As seen in models of photoevaporation and core-powered mass-loss in Figure \ref{fig:SincPrad}, the radius gap extends to high incident flux, whereupon it opens into a large sparsity of planets referred to as the ``Neptune-desert''. When $\alpha$ and $\beta$ are being extracted from the 3D data (as in Section \ref{sec:GradientCalc}), this wedge-shaped feature frequently causes the extracted value of $\beta$ to be diminished. However, as incident flux is strongly correlated with stellar mass, if we move to lower mass stars, the Neptune-desert becomes less prevalent in the data. As a result, the algorithm can determine $\beta$ more accurately.

\begin{figure} 
	\includegraphics[width=1.0\columnwidth]{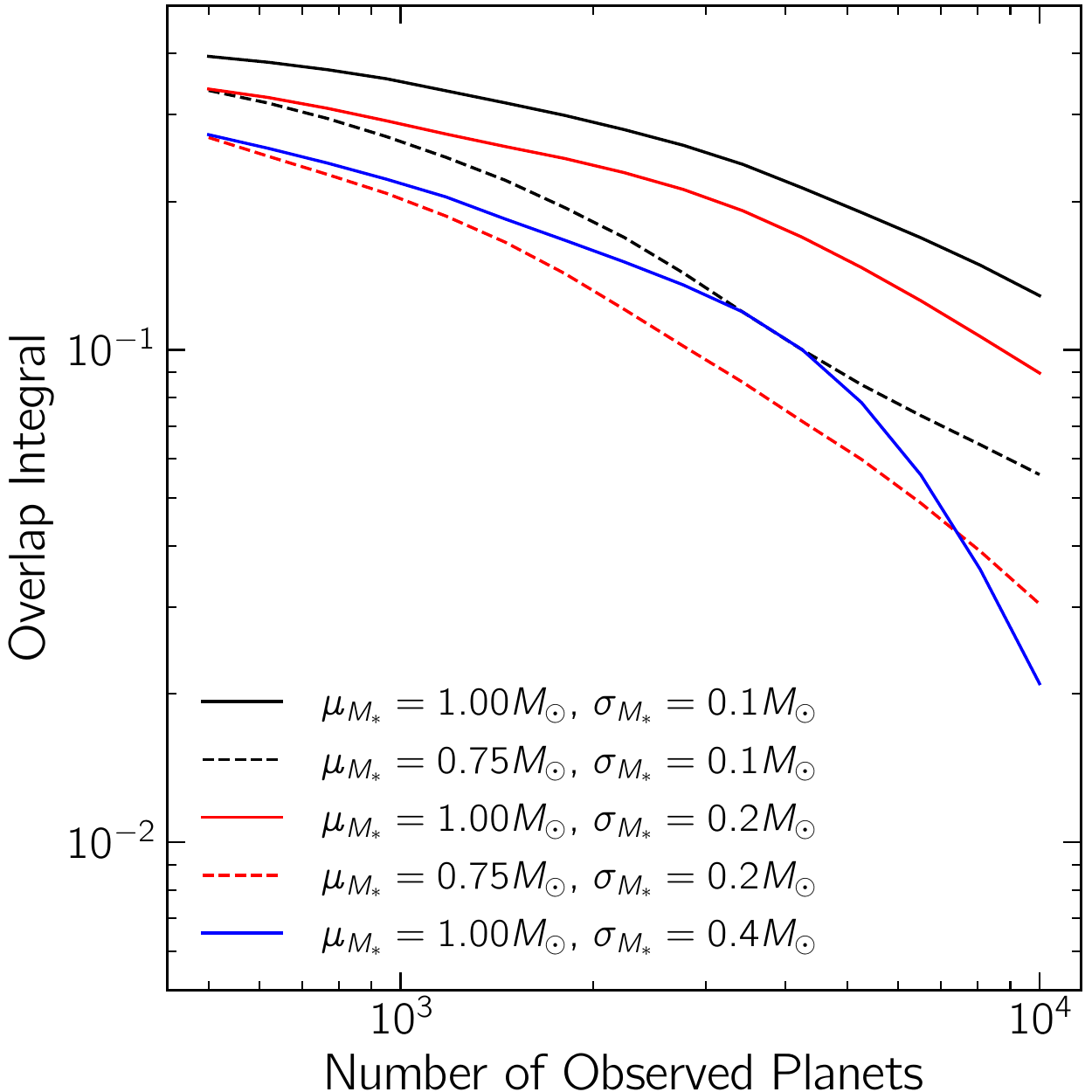}
    \caption{Same as Figures \protect{\ref{fig:OverlapNominal}} and \protect{\ref{fig:OverlapUncFP}}, whereas here the overlap integral is calculated for different stellar mass distribution means $\mu_{M_*}$ and standard deviations $\sigma_{M_*}$. Dashed lines represent surveys where the stellar mass distribution is centred at lower masses, which has the effect of reducing the overlap integral for the same distribution width. A survey with a distribution centred at $1.0M_\odot$ and large width of $0.4M_\odot$ is shown in blue for comparison.}
    \label{fig:OverlapSmassDist} 
\end{figure}

To demonstrate this, Figure \ref{fig:OverlapSmassDist} shows the overlap integrals of a selection of our baseline surveys in Figure \ref{fig:OverlapNominal} compared with surveys in which the stellar mass distribution is shifted to a Gaussian function centred at $0.75M_\odot$. We clearly see that shifting the distributions to lower stellar masses reduces the overlap integral and thus improves ability to distinguish photoevaporation from core-powered mass-loss. For a survey with stellar mass distribution centred at $0.75M_\odot$ and width $0.2M_\odot$, photoevaporation can be distinguished from core-powered mass-loss with a survey of $\gtrsim 3000$ planets $90\%$ of the time. We also include a survey centred at $1.0M_\odot$ with a large width of $0.4M_\odot$ to show that one can achieve the same, if not a better, result if one surveys a narrow mass range of stars but biased to lower masses. 

\subsection{Changing Model Parameters} \label{sec:modelparams}

\subsubsection{Photoevaporation}
As discussed in Section \ref{sec:PE-Predictions}, the predicted value of $\beta$ can be altered if one assumes different model parameters for the high-energy stellar evolution in the photoevaporation model. As a result, we intentionally designed the model such that the predicted value of $\beta$ was as small in magnitude (and therefore closest to the value predicted by core-powered mass-loss) as possible, whilst still being physically justified. This results in shifting the posterior of photoevaporation closer to that of core-powered mass-loss, and hence systematically increases the overlap integral. By presenting a conservative model for photoevaporation, we make the task of performing this statistical analysis more challenging, but can state more securely that if this were done on real data, any definitive result cannot be negated by large changes in the models of photoevaporation.

\begin{figure} 
	\includegraphics[width=1.0\columnwidth]{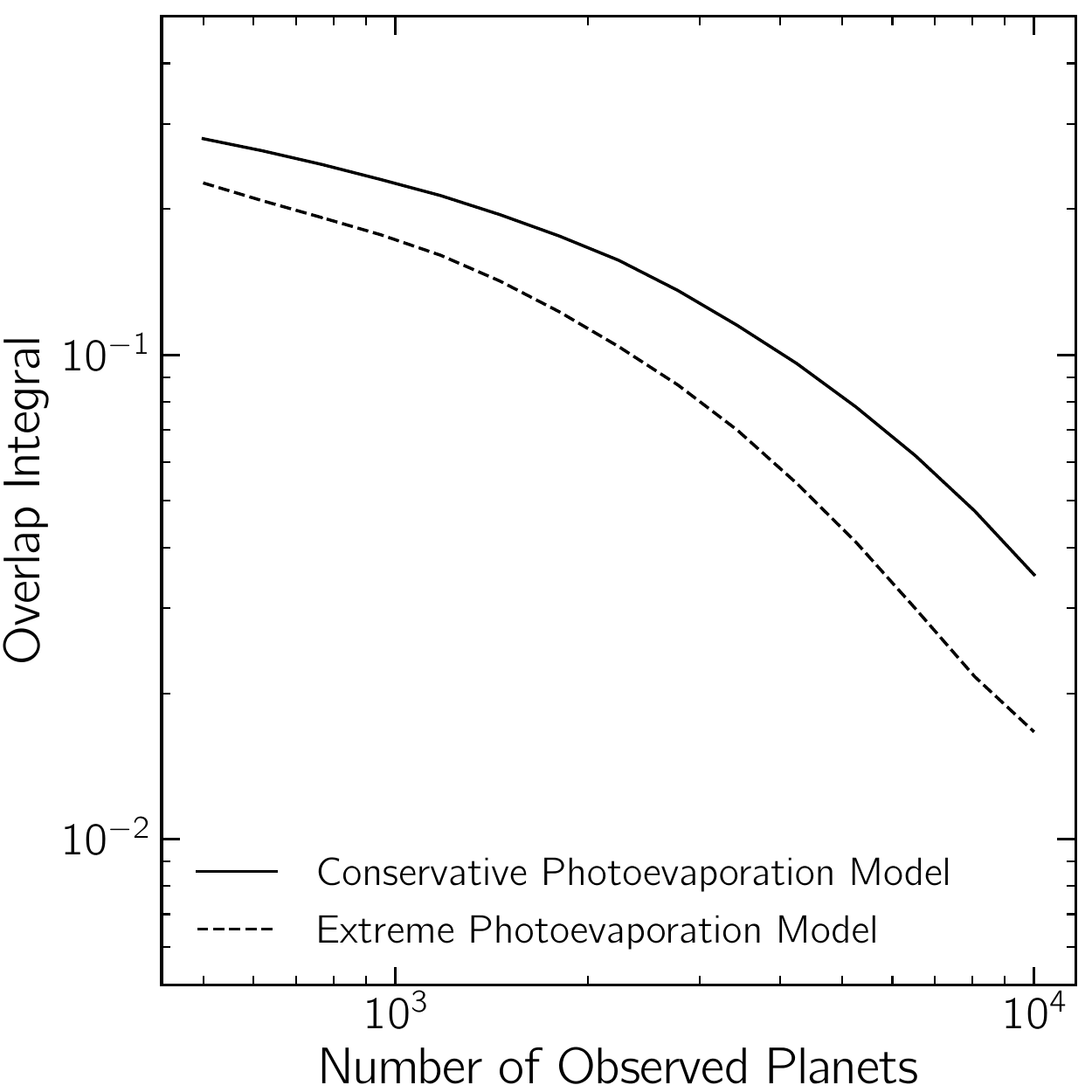}
    \caption{Same as Figures \protect{\ref{fig:OverlapNominal}}, \protect{\ref{fig:OverlapUncFP}} and \protect{\ref{fig:OverlapSmassDist}}, whereas here the overlap integral is calculated under different assumptions for the photoevaporation model. Both surveys have a stellar mass distribution centred at $1.0M_\odot$ and width of $0.3M_\odot$. The conservative model (i.e. as presented in Section \protect{\ref{sec:PE}}) is shown in the solid black line, whereas a more extreme model with a larger magnitude of predicted $\beta$ is shown in the dashed line.}
    \label{fig:OverlapExtreme} 
\end{figure}

With that in mind however, if photoevaporation is the driving mechanism of atmospheric mass-loss, it may in reality result in a larger {\jr{magnitude of}} $\beta$. To demonstrate the effect this would have, Figure \ref{fig:OverlapExtreme} shows the overlap integral calculated for two surveys: one in which the photoevaporation model is the conservative one presented in Section \ref{sec:PE} {\jr{with $\beta=-0.17$}}, whilst the other is a more extreme model, where $\beta = -0.23$ is the new predicted value. To do this, we change the stellar mass scalings of X-ray to bolometric luminosity $L_\text{XUV} / L_\text{bol}$ to be:
\begin{equation} \label{eq:LxLbolNew}
    \frac{L_\text{XUV}}{L_\text{bol}}=\begin{cases}
    \bigg( \frac{L_\text{XUV}}{L_\text{bol}} \bigg)_{\text{sat}} \; \bigg( \frac{M_*}{M_\odot} \bigg)^{-1.0} & \text{for } t < t_\text{sat}, \\
    \bigg( \frac{L_\text{XUV}}{L_\text{bol}} \bigg)_{\text{sat}} \; \bigg( \frac{M_*}{M_\odot} \bigg)^{-1.0} \; \bigg ( \frac{t}{t_\text{sat}} \bigg)^{-1-a_0}  & \text{for } t \geq t_\text{sat}.
    \end{cases}
\end{equation}
Note that before (as in Eq. \ref{eq:LxLbol}), the index on stellar mass was $-0.5$ as opposed to $-1.0$ in this case. As is clear in Figure \ref{fig:OverlapExtreme}, the overlap decreases assuming this form of photoevaporation is taking place, meaning that distinguishing between it and core-powered mass-loss could potentially occur for a smaller survey.

\subsubsection{Core-Powered Mass-Loss}
Unlike photoevaporation, the predicted values of $\alpha$ and $\beta$ for core-powered mass-loss are more certain. One possible avenue to consider however, is the metallicity trend of host stars \citep[][]{Gupta2020}. Let us suppose that host star metallicities and masses are positively correlated: $Z_* \propto M_*^a$ where $a>0$ is some constant. In addition, following \citet{Gupta2020}, let us assume that the metallicities of primordial planet atmospheres follow that of their host stars such that [Fe/H] $\simeq [Z_*/Z_\odot]$, which is not unreasonable because metal-rich stars are expected to have metal-rich discs. It then follows that more massive stars will host planets with metal enhanced atmospheres. Furthermore, assuming that metal-rich atmospheres have higher opacities ($\kappa$) such that $\kappa \propto Z_*$, implies that more massive or metal-rich stars are likely to host planets with higher opacity atmospheres. Consequently, more massive stars are likely to host planets that are inflated in sizes for their age because of their longer cooling timescales.

For planets hosting significant H/He atmospheres i.e. sub-Neptunes, this will result in a positive slope in their size as a function of stellar mass. Recall however, that $\alpha$ and $\beta$ are defined as slopes in the maximum size of stripped cores i.e. super-Earths, as a function of $S$ and $M_*$. Therefore, a slope in the sizes of sub-Neptunes will not effect either parameter. This is true in practise as well, we find that the extracted values of $\alpha$ and $\beta$ are consistent and independent of metallicity trends. Figure \ref{fig:MethodLimitation} demonstrates that the gradient extraction method, when applied to core-powered mass-loss, follows the trends in the sizes of super-Earths. If one were to incorporate metallicity trends, the sub-Neptune slope would change but this would only negligibly effect the slope of super-Earths and hence the extracted values of $\alpha$ and $\beta$. Note that since both models incorporate atmospheric cooling, this conclusion also likely holds for photoevaporation but is less certain due to the critical role of metals in the absorbing of high-energy radiation and in regulating cooling when in the escaping layers of the atmosphere \citep[][]{Owen2018}. 

\subsection{Results for Existing Surveys}
\begin{figure*} 
	\includegraphics[width=2.0\columnwidth]{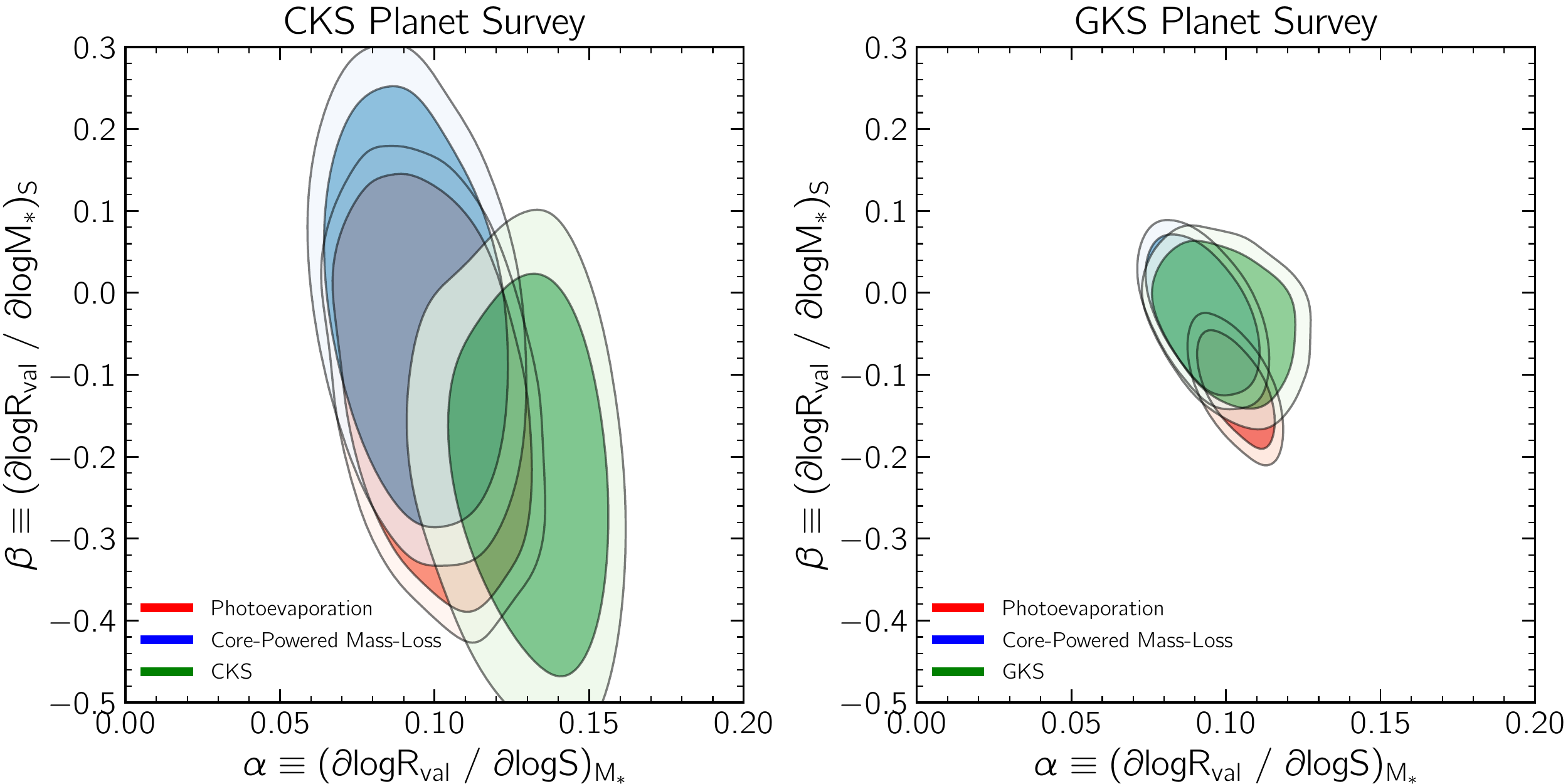}
    \caption{Posteriors for $\alpha$ and $\beta$ are shown in green for the CKS data, which has 921 planets \protect{\citep{Fulton2017,Fulton2018}} (left) and the GKS survey with 3760 planets \protect{\citep{Berger2020}} (right), with contours representing $1\sigma$ and $2\sigma$ confidence levels. Red and blue contours represent synthetic surveys using the photoevaporation model and core-powered mass-loss model respectively. The synthetic surveys match the number of detected planets and assume a stellar mass distribution centred at $1M_\odot$ with widths of $0.15M_\odot$ and $0.2M_\odot$ for CKS and GKS respectively. Note that both assume measurement uncertainties in incident flux, stellar mass and planet size as ($8\%$,$5\%$,$5\%$) and both adopt the completeness map from the CKS survey.}
    \label{fig:CKSvsGKS} 
\end{figure*}

To test the methodology, we perform the analysis on two surveys which both derive from \textit{Kepler} data. The CKS dataset \citep{CKSI-Petigura2017,Fulton2017,Fulton2018} is a sample of 921 planets with high purity {\jr{(i.e. low levels of false positive contamination with further cuts on effective temperature, impact parameter and removal of giant stars, see \citet{Fulton2017} for details)}}. We model it's stellar mass distribution as a Gaussian function centred at $1M_\odot$ and width of $0.15M_\odot$ \citep{Rogers2021}. The GKS survey on the other hand \citep{Berger2020} has a sample of 3760 planets but includes planet candidates which are (as of yet) not confirmed, even after the appropriate cuts. As a result, the level of false positive contamination may be significantly larger than that of CKS. This survey has a slightly broader stellar mass distribution, which we model as being Gaussian with centre of $1M_\odot$ and width of $0.2M_\odot$. We construct synthetic transit survey with photoevaporation and core-powered mass-loss models assuming the same number of planets as the individual surveys and the appropriate stellar mass distributions. Note that, as shown in Section \ref{sec:FP}, false positive contamination has little effect on determining $\alpha$ and $\beta$, hence we do not include them in the synthetic surveys. In addition, both surveys benefit from \textit{Gaia} stellar mass constraints and thus we model the measurement uncertainty in incident flux, stellar mass and planet size as \{$8\%$, $5\%$, $5\%$\} for both synthetic surveys. One caveat is that the completeness map from the CKS data is used for both synthetic surveys. However, as both data-sets derive from the same underlying planet population and were observed using the same telescope, this is a fair approximation to make.

Figure \ref{fig:CKSvsGKS} shows the posteriors for the CKS data (left) and GKS data (right) alongside the appropriate synthetic surveys. As expected, the posterior for GKS is significantly smaller, owing to its larger size and wider stellar mass distribution. {\jr{While we measure $\alpha = 0.13^{+0.03}_{-0.05}$ and $\beta=-0.21^{+0.33}_{-0.39}$ for the CKS data, we find $\alpha = 0.10^{+0.03}_{-0.02}$ and $\beta=-0.03^{+0.10}_{-0.12}$ for the GKS data, with all uncertainties quoted as $2\sigma$.}} Both surveys are consistent with each other, and also consistent with synthetic surveys for photoevaporation and core-powered mass-loss. As a result, we can not conclude the data favours one model over the other at this stage.

\section{Discussion}\label{sec:Discussion}
In this work we have analysed the radius gap in 3D parameter space of incident flux, stellar mass and planet size. We have shown that it is suitable for model comparisons, particularly because $\beta = (\partial \log R_\text{val} / \partial \log M_*)_{S}$ is predicted to be significantly different between the two models. Comparing this to model comparisons made in the $M_*$-$R_\text{p}$ plane, in which the slope of the radius valley is fraught with degeneracies of $\alpha$, $\beta$ as well as any scaling between core mass and stellar mass, it follows that the adopted 3D method far outperforms the 2D equivalent.

\subsection{The Perfect Survey}
From the results presented, the hypothetical perfect survey to understand which of photoevaporation or core-powered mass-loss is the driving mechanism in exoplanet evolution would have the following traits:
\begin{itemize}
    \setlength\itemsep{1em}
    \item Biased towards host stellar masses $\leq 1.0M_\odot$, but with a typical spread of $\geq 0.2 M_\odot$.
    
    \item Emphasis on reducing measurement uncertainty, with typical percentage errors $\leq8\%$ for incident flux, $\leq5\%$ for stellar mass and $\leq5\%$ for planet radius.
    
    \item A false positive contamination of $\leq10\%$.
\end{itemize}
In this scenario, core-powered mass-loss may be distinguished from a conservative model of photoevaporation $90\%$ of the time with a survey of $\gtrsim 3000$ planets. However, given that the majority of currently observed planets are centred around $\sim1.0M_\odot$, a more realistic survey which includes these may be able to achieve this goal with $\gtrsim 5000$ planets.

For future surveys, we propose the following process. Determine $\alpha$ and $\beta$ in a similar manner to that of Section \ref{sec:GradientCalc} and repeat by re-sampling planet measurements within the associated error range. The priority is that the adopted method must be robust i.e. it consistently extracts $\alpha$ and $\beta$ for any resample of the data and the values appear to be correct by-eye. Then, a hypothesis test should be performed, with the null hypothesis being that $\beta=0.0$ (i.e. consistent with core-powered mass-loss). This can then be rejected in favour of the alternative hypothesis $\beta\neq0.0$ (i.e. favouring photoevaporation) if the data suggests so, to some significance level. Note that the posteriors presented in this work from synthetic models will differ from real surveys, since in our work we are determining the uncertainty in $\alpha$ and $\beta$ by performing $10,000$ individual surveys to understand the Poisson error. For real data however, uncertainties in $\alpha$ and $\beta$ are determined by resampling over the individual planet measurements. 

\subsection{The Effect of Survey Completeness}
\begin{figure} 
	\includegraphics[width=1.0\columnwidth]{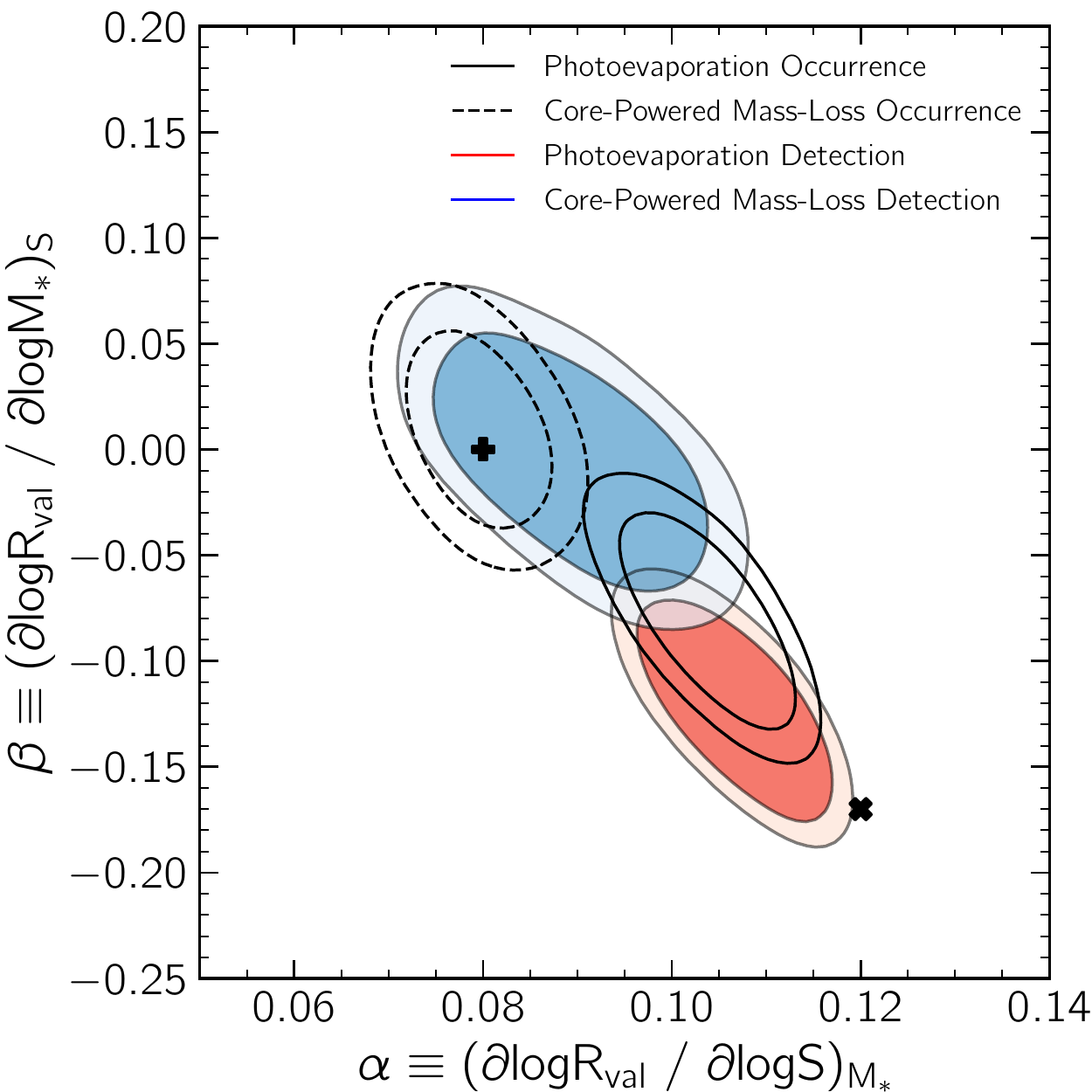}
    \caption{Posteriors for $\alpha$ and $\beta$ are shown for photoevaporation and core-powered mass-loss for a survey with host stellar mass distribution centred at $1.0M_\odot$ with a width of $0.3M_\odot$. Coloured contours represent the synthetic survey measurements for photoevaporation (red) and core-powered mass-loss (blue) with observational bias, selection effects and measurement uncertainty incorporated. Black contours represent the posteriors when $\alpha$ and $\beta$ are extracted from the underlying planet distributions provided by the models, referred to as the occurrence, with solid lines representing photoevaporation and dashed lines for core-powered mass-loss. Black markers represent the theoretical predictions for each model.}
    \label{fig:OccVsDet} 
\end{figure}

All surveys, including transit surveys, have inherent bias folded into the data. In this case, the bias arises due to the geometric probability of transit and also pipeline efficiency for a given exoplanet. As a result, detections are biased to be at short orbital periods (large incident flux) and large planet radii. Although this is unavoidable, it is worth noting that this does have an effect on the values of $\alpha$ and $\beta$ that can be extracted from data. Figure \ref{fig:OccVsDet} shows posteriors in colour for photoevaporation and core-powered mass-loss for a survey of 5000 planets, measurement uncertainty in $(S, M_*, R_\text{p})$ of $(8\%, 5\%, 5\%)$ and a stellar mass distribution of a Gaussian function centred at $1.0M_\odot$ with a width of $0.3M_\odot$. Recall that in order for a synthetic transit survey to be produced, we bias our underlying distribution with the CKS completeness map i.e. all radius gap analysis thus far has been performed in detection space (see Section \ref{sec:TransitSurveys}). To demonstrate the effect of bias, we also show the posteriors in solid black and dashed contours for $\alpha$ and $\beta$ extracted from the underlying distributions i.e. no bias and no measurement uncertainty, labelled as the occurrence. The effect is clearly seen for core-powered mass-loss: the posterior for the occurrence is centred correctly on the theoretically predicted value, whereas the posterior for the biased and noisy data is larger and shifted to larger values of $\alpha$ and smaller values of $\beta$. The increase in posterior size is due to measurement uncertainty (also demonstrated in Figure \ref{fig:OverlapUncFP}), whereas the shift in position arises due to survey bias. As survey incompleteness is not uniform i.e. it preferentially biases towards larger $S$ and larger $R_\text{p}$, it introduces biases in the slope of the radius gap. This is similar to photoevaporation (left panel of Figure \ref{fig:OccVsDet}) in that the posterior of the biased survey is shifted to larger values of $\alpha$ and smaller values of $\beta$. The difference however, is that the posterior distribution in fact gets closer to the theoretically predicted values. This is due to an unrelated but inherent limitation of the gradient extraction method that is further discussed in Section \ref{Sec:Limitations}.

Despite the introduction of systematic errors and as further argued in \cite{Rogers2021}, it is important to do this analysis in detection space as opposed to performing completeness corrections and performing analysis in occurrence space. This is because planets detected with low completeness (i.e. low probability of being detected such as those with large orbital periods), are given large weights and significantly increase the inferred occurrence in that region. Instead, statistical analysis should be done in detection space, whilst predictions from the models should be calculated by forward modelling planets through a synthetic transit survey so that they can be compared. Note that this then requires completeness maps to be calculated for a given survey so that they can be used with the theoretical models.

\subsection{Limitations of the Statistical Analysis} \label{Sec:Limitations}
One of the main drawbacks of this method is its partial ineffectiveness in extracting the correct value of $\beta$ from the photoevaporation models. This is discussed in Section \ref{sec:GradientCalc} and can be seen in Figures \ref{fig:MethodLimitation}, \ref{fig:Summary} and in the left side of \ref{fig:OccVsDet}, in which the posterior distributions for photoevaporation are inconsistent with the theoretical predictions. The cause of this limitation originates from a subtle difference in the definition of the radius gap between the models and the data. In the analytic models, the valley slope is defined by locating the size of the largest super-Earths as a function of either orbital period, incident flux or stellar mass (see Sections \ref{sec:PE-Predictions} and \ref{sec:CPML-Predictions}). The adopted computational method of extracting these slopes however is to find a plane in the three-dimensional space of incident flux, stellar mass and planet size which intersects with the fewest number of planets and tends to bisect the two populations of super-Earths and sub-Neptunes (see Section \ref{sec:GradientCalc}). Whilst these two definitions are similar, the theoretical value is consistently steeper for photoevaporation. As a result, the value extracted is shallower in $\beta$ than the predicted values. Overall, this issue stems from the fact that characterising the radius gap is an ill-posed statistical problem. {\jr{Note that this also holds true for other methods of radius gap characterisation, such as \textit{gapfit} from \citet{Loyd2020}.}} When performing this analysis on future surveys, other, potentially more sophisticated statistical frameworks may be set up in order to extract $\alpha$ and $\beta$. This new framework should, as done in this work, be implemented on modelled data as well as surveys such that observational data is not directly compared to theoretical predictions, but synthetic surveys instead. For now though, this is beyond the scope of this work.

\subsection{Independence of Core Mass Scaling} \label{sec:CoreMassScaling}

As discussed in Section \ref{sec:alpha&beta}, analysing the gap in 3D is independent of scalings of core mass with stellar mass. This is contrary to the $M_*$-$R_\text{p}$ plane, in which scalings can alter the slope of the radius gap \citep[e.g.][]{Wu2019} and further increase the degeneracy in its physical origin (i.e. Figure \ref{fig:SmassPrad}). Whilst in theory this claim is true when applied to the analytic models, in practise the applied method of gradient extraction suffers from small systematic shifts in measured $\beta$ when scalings of core mass to stellar mass are introduced. This originates from the same issues discussed in Section \ref{Sec:Limitations}, in which the theoretically predicted radius gap is not what is extracted from the survey data. This is therefore further evidence that refining the statistical method of radius gap detection to be closer to the theoretical definitions will improve future analysis.

\subsection{A Route Forward}
Eventually, determining which of photoevaporation or core-powered mass-loss determines exoplanet evolution will become a multi-faceted endeavour. In this work, we have presented a new avenue to explore, in which the radius gap is extended to 3D space and mass-loss parameters $\alpha$ and $\beta$ are extracted. This method should be combined with other techniques capable of differentiating between the models, such as radius gap evolution \citep[e.g.][]{Berger2020,Sandoval2020} and direct observation of atmospheric mass-loss through transit spectroscopy \citep[e.g.][]{Gupta2021} of Lyman-$\alpha$ \citep[e.g.][]{VidalMadjar2003,Ehrenreich2015}, H$\alpha$ \citep[e.g.][]{Yan2018}, He 1083 nm \citep[e.g.][]{Spake2018,Ninan2020} or other metal lines \citep[e.g.][]{VidalMadjar2004}.

In addition, further observational work is needed in order to constrain the X-ray evolution of stars, particularly how X-ray exposure changes with stellar mass \citep[e.g.][]{McDonald2019}, {\jr{in addition to the level of intrinsic scatter of X-ray exposure for a given stellar mass \citep[e.g.][]{Tu2015}.}} This has a direct effect on the predicted value of $\beta$ for photoevaporation, hence further constraints will aid in making model predictions. 

Finally, further theoretical work is needed to construct a combined model of photoevaporation and core-powered mass-loss. A major assumption in this work is that the physics of the two models are mutually exclusive. However, in reality, it is likely that both mechanisms have a part to play in exoplanet evolution. Constructing a joint model which encapsulates the physics of both mechanisms is therefore part of the planned future work.

Once knowledge is gained as to which mechanism dominates the evolution of close-in exoplanets, one can exploit the evolutionary process to `rewind the clock' and infer underlying demographic distributions that are  hidden in standard surveys. This was first performed with photoevaporation by \citet{Wu2019} and then again by \citet{Rogers2021}, in which constraints were placed on the core mass distribution, initial atmospheric mass fraction distribution and core composition distribution. \citet{Gupta2020} performed a similar analysis to infer underlying planet and stellar parameters for the core-powered mass-loss mechanism.

\section{Conclusions}\label{sec:conclusions}
In this work, we have presented a new method of differentiating between the atmospheric mass-loss models of EUV/X-ray photoevaporation \citep[e.g.][]{Owen2013,Owen2017} and core-powered mass-loss \citep[e.g.][]{Ginzburg2016,Gupta2019} using exoplanet demographics. This involves analysing the radius gap in demographic data in the 3D parameter space of incident flux $S$, host stellar mass $M_*$ and planet size $R_\text{p}$. We define two new parameters, $\alpha = (\partial \log R_\text{val} / \partial \log S)_{M_*}$ and $\beta = (\partial \log R_\text{val} / \partial \log M_*)_{S}$ and a statistical method of extracting them from transit data. We show that performing model comparisons in this 3D space is superior to considering the gap in 2D (such as the $M_*$-$R_\text{p}$ plane) as it avoids degeneracies of $\alpha$, $\beta$ as well as scalings of core mass with stellar mass \citep[e.g.][]{Wu2019,Gupta2020}. We perform synthetic transit surveys for both models of varying size and quality in order to determine the best route forward to determine which of the two models best describes the observed exoplanet population. Our main conclusions are as follows:

\begin{itemize}
    \setlength\itemsep{1em}
    \item The predicted values of $\alpha$ are $\sim0.08$ and $\sim0.12$ for core-powered mass-loss and photoevaporation respectively. The predicted values of $\beta$ are $\sim0.00$ and $\sim-0.17$ respectively. This difference arises because the energy source for photoevaporation originates from the high energy flux exposure from the star. At fixed incident bolometric flux, this has been shown to decrease with host stellar mass \citep{McDonald2019}, resulting in a negative value of $\beta$. On the other hand, the energy source for the core-powered mass-loss is a combination of core luminosity and stellar bolometric luminosity, implying $\beta$ is predicted to be zero, from its definition.
    
    \item We predict that transit surveys with $\gtrsim 5000$ planets will have the statistical power necessary to extract $\alpha$ and $\beta$ such that one can begin to determine which model best describes the data. This applies to a transit survey with a stellar mass distribution centred at $1.0M_\odot$ and standard deviation of $0.3M_\odot$, as well as typical measurement uncertainties of $\leq8\%$ for incident flux, $\leq5\%$ for stellar mass and $\leq5\%$ for planet radius.
    
    \item We find that false positive contamination of up to $10\%$ does not effect the ability to perform this analysis on a given transit survey, whereas reducing the measurement uncertainty has a major positive impact. Additionally, emphasis should be put on increasing the width of the host stellar mass distribution, as this provides greater statistical power to determine $\beta$. Biasing towards host stellar masses $<1M_\odot$ also reduces the negative impact of the Neptune-desert on the analysis.
    
    \item This analysis was performed on two existing surveys: the CKS data-set \citep{Fulton2018} and the GKS data-set \citep{Berger2020}. We find that the extracted values for $\alpha$ and $\beta$ from both surveys are consistent with each other and also consistent with both photoevaporation and core-powered mass-loss synthetic surveys. 
    
    \item Additionally, we show that performing model comparisons in the $M_*$-$R_\text{p}$ plane is fraught with degeneracies between $\alpha$, $\beta$ and the presence of any stellar mass-core mass scalings. This motivates the analysis to be performed in 3D $S$-$M_*$-$R_\text{p}$ space, as performed in this work, in order to break these degeneracies. 
    
    \item {\jr{As argued in \citet{Rogers2021}, we further stress that comparisons between data and models should be performed in detection space as opposed to completeness-corrected occurrence space, which can bias results due to planets in low-completeness regions being given large weights.}} 
    
    \item Finally, once the dominant evolutionary mechanism of close-in exoplanets is determined, the associated model can be exploited to `rewind the clock' and infer demographic distributions that are hidden in surveys such as the core mass distribution, initial atmospheric mass fraction distribution and core composition distribution \citep{Wu2019,Gupta2020,Rogers2021}.

\end{itemize}

Looking forward to new surveys, the introduction of data from missions such as \textit{K2} \citep[e.g.][]{Howell2014, Zink2020} and \textit{TESS} \citep[e.g.][]{Ricker2015} are encouraging for this avenue of model comparison. In addition, future survey missions such as \textit{PLATO} \citep{PLATO2014} will provide observations of thousands of additional exoplanets. {\jr{In light of our results, we would encourage future demographic missions to put emphasis on sampling a broad range of host stellar masses and utilising high accuracy planetary radii and age determination (e.g. asteroseismology) for planets near or inside the radius gap.}} For now however, the fundamental question still remains: what caused close-in exoplanets to lose their atmospheres? To answer this, we will require further work to be done on both observational and theoretical sides of the field. Besides the demand for larger surveys, further constraints are needed on the X-ray evolution of young stars, which will aid in the understanding of the photoevaporation models. Additionally, pursuing other model comparison avenues such as transit spectroscopy and radius gap time evolution will aid in answering this question. {\jr{Finally, theoretical work is needed to further develop both evolutionary models. Both of the adopted formalisms include a series of theoretical assumptions that require further investigation. In addition it is unclear how photoevaporation and core-powered mass-loss compete and which, if any, dominates under varying conditions.}}

\section*{Acknowledgements}
{\jr{We kindly thank the anonymous reviewer, as well as Travis Berger, Parke Loyd, Erik Petigura, James Sikora and Yanqin Wu for comments and discussion that improved the paper.}} We are grateful to the CKS and GKS teams for making their results public, which allowed this analysis to be performed. JGR is supported by a 2017 Royal Society Grant for Research Fellows and a 2019 Warner Prize. AG is supported by the Future Investigators in NASA Earth and Space Science and Technology (FINESST) grant 80NSSC20K1372. JEO is supported by a Royal Society University Research Fellowship. This work was supported by a 2020 Royal Society Enchantment Award and this project has received funding from the European Research Council (ERC) under the European Union’s Horizon 2020 research and innovation programme (Grant agreement No. 853022, PEVAP). HES gratefully acknowledges support from NASA under grant number 80NSSC21K0392 issued through the Exoplanet Research Program. This work was performed using the DiRAC Data Intensive service at Leicester, operated by the University of Leicester IT Services, which forms part of the STFC DiRAC HPC Facility (www.dirac.ac.uk). The equipment was funded by BEIS capital funding via STFC capital grants ST/K000373/1 and ST/R002363/1 and STFC DiRAC Operations grant ST/R001014/1. DiRAC is part of the National e-Infrastructure. JGR and JEO are grateful to hospitality from UCLA where this work initiated. 

\section*{Data Availability}
The data underpinning the measured posteriors of $\alpha$ and $\beta$ for a range of synthetic surveys can be found on Zenodo at \href{https://doi.org/10.5281/zenodo.4738492}{10.5281/zenodo.4738492}.




\bibliographystyle{mnras}
\bibliography{references} 

\bsp	
\label{lastpage}
\end{document}